\documentclass[reprint,superscriptaddress,amssymb,amsmath,aps,prl,longbibliography]{revtex4-2}
\usepackage{setspace}
\usepackage{graphicx}% Include figure files
\usepackage{dcolumn}% Align table columns on decimal point
\usepackage{bm}% bold math
\usepackage{siunitx}
\usepackage{chemformula}
\usepackage{comment}
\usepackage[resetlabels,labeled]{multibib}
\usepackage{placeins}
\newcites{Main}{blah1}
\newcites{Methods}{blah2}

\begin{document}

%\preprint{APS/123-QED}

\title{Non-destructive optical readout of a superconducting qubit}% Force line breaks with \\
\author{R.~D.~Delaney}
\email{robert.delaney@colorado.edu}
\affiliation{JILA, National Institute of Standards and Technology and the University of Colorado, Boulder, Colorado 80309, USA}
\affiliation{Department of Physics, University of Colorado, Boulder, Colorado 80309, USA}
\author{M.~D.~Urmey}
\affiliation{JILA, National Institute of Standards and Technology and the University of Colorado, Boulder, Colorado 80309, USA}
\affiliation{Department of Physics, University of Colorado, Boulder, Colorado 80309, USA}
\author{S.~Mittal}
\affiliation{JILA, National Institute of Standards and Technology and the University of Colorado, Boulder, Colorado 80309, USA}
\affiliation{Department of Physics, University of Colorado, Boulder, Colorado 80309, USA}
\author{B.~M.~Brubaker}
\affiliation{JILA, National Institute of Standards and Technology and the University of Colorado, Boulder, Colorado 80309, USA}
\affiliation{Department of Physics, University of Colorado, Boulder, Colorado 80309, USA}
\author{J.~M.~Kindem}
\affiliation{JILA, National Institute of Standards and Technology and the University of Colorado, Boulder, Colorado 80309, USA}
\affiliation{Department of Physics, University of Colorado, Boulder, Colorado 80309, USA}
\author{P.~S.~Burns}
\affiliation{JILA, National Institute of Standards and Technology and the University of Colorado, Boulder, Colorado 80309, USA}
\affiliation{Department of Physics, University of Colorado, Boulder, Colorado 80309, USA}
\author{C.~A.~Regal}
\affiliation{JILA, National Institute of Standards and Technology and the University of Colorado, Boulder, Colorado 80309, USA}
\affiliation{Department of Physics, University of Colorado, Boulder, Colorado 80309, USA}

\author{K.~W.~Lehnert}
\affiliation{JILA, National Institute of Standards and Technology and the University of Colorado, Boulder, Colorado 80309, USA}
\affiliation{Department of Physics, University of Colorado, Boulder, Colorado 80309, USA}
\affiliation{National Institute of Standards and Technology, Boulder, Colorado 80305, USA}

\pacs{Valid PACS appear here}% PACS, the Physics and Astronomy
\maketitle
\textbf{Entangling superconducting quantum processors via light would enable new means of secure communication and distributed quantum computing \cite{arute2019quantum}.  
However, transducing quantum signals between these disparate regimes of the electromagnetic spectrum remains an outstanding goal \cite{ hisatomi2016bidirectional,han2018coherent,higginbotham2018harnessing,bartholomew2020chip,stockill2021ultra-low-noise,sahu2021quantum-enabled}, 
and interfacing superconducting qubits with electro-optic transducers presents significant challenges due to the deleterious effects of optical photons on superconductors \cite{barends2011minimizing,mirhosseini2020superconducting}.  Moreover, many remote entanglement protocols \cite{campbell2008measurement, kalb2017entanglement, zhong2020proposal, kurpiers2019quantum} require multiple qubit gates both preceding and following the upconversion of the quantum state, and thus an ideal transducer should leave the state of the qubit unchanged: more precisely, the backaction \cite{hatridge2013quantum} from the transducer on the qubit should be minimal.
Here we demonstrate non-destructive optical readout of a superconducting transmon qubit via a continuously operated electro-optic transducer.  The modular nature of the transducer and circuit QED system used in this work enable complete isolation of the qubit from optical photons, and the backaction on the qubit from the transducer is less than that imparted by thermal radiation from the environment. 
Moderate improvements in transducer bandwidth and added noise will enable us to leverage the full suite of tools available in circuit QED to demonstrate transduction of non-classical signals from a superconducting qubit to the optical domain.}

Superconducting quantum computers using arrays of transmon qubits are a leading platform for scalable quantum computation \cite{arute2019quantum}.  However, these devices must operate at temperatures below $100$~mK, with quantum information encoded in microwave fields that would be corrupted by thermal noise if transmitted at ambient temperature.  In contrast, optical quantum networks are a well-established technology for the transmission of quantum states over long distances, and do not require low temperatures \cite{inagaki2013entanglement,ursin2007entanglement, yin2017satellite}.  Thus, a quantum-enabled electro-optic transducer linking the microwave and optical domains would greatly expand the capabilities of quantum information science.

\begin{figure*}
    \centering
    \includegraphics{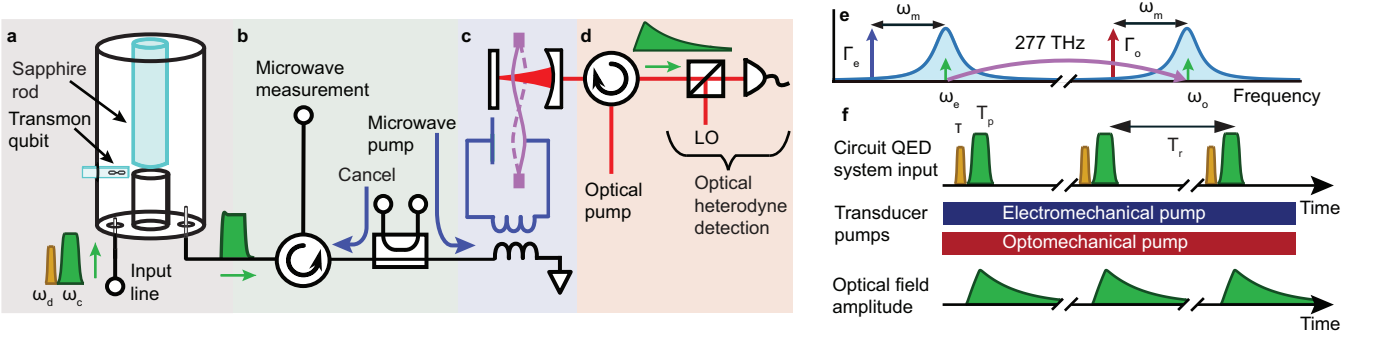}
    \caption{\textbf{Apparatus for optically mediated readout of a superconducting qubit} (a) Circuit QED system consisting of a transmon qubit dispersively coupled to a 3D coaxial quarter-wave cavity resonator.   A translatable sapphire rod tunes the frequency of the cavity \textit{in situ}.  Qubit preparation pulses with frequency $\omega_\textrm{d}$ (gold), followed by readout pulses with frequency $\omega_\textrm{c}$ (green), are injected through the circuit QED system's input line.  (b) The directional coupler is used to apply the microwave pump to the transducer, while the circulators enable microwave heterodyne measurement of reflected signals.  Isolation between the circuit QED system and the transducer is achieved using circulators, along with the directional coupler for interferometric cancellation.  (c) The electro-optic transducer consists of an optical cavity and a flip-chip microwave LC circuit resonator simultaneously coupled to a single mode of a high-quality factor silicon nitride membrane with frequency $\omega_\textrm{m}/2\pi= 1.45$~MHz. (d) Optical pump and heterodyne detection scheme. (e)  Transducer pumps and experimental readout signals represented in the frequency domain.  The electro(opto)-mechanical damping rates $\Gamma_\textrm{e}$ ($\Gamma_\textrm{o}$) are controlled by the strength of the respective pumps, both red-detuned by $\omega_\textrm{m}$ to transduce microwave signals to the optical domain. (f) A pulse timing diagram illustrates the qubit (gold) and readout (green) pulses, with the experiment conducted with a repetition time of $T_\textrm{r} = 0.4-2$~ms and steady-state pumps. The upconverted readout pulse is filtered by the transducer's frequency response (bottom row).}
    \label{fig:fig1}
\end{figure*}

The pursuit of an optical quantum network of superconducting qubits has given rise to a rich research field searching for efficient, low-noise electro-optic transduction techniques \cite{hisatomi2016bidirectional, han2018coherent, higginbotham2018harnessing, stockill2021ultra-low-noise, mirhosseini2020superconducting, bartholomew2020chip, sahu2021quantum-enabled}. 
Although electro-optic elements have been demonstrated as useful tools for delivering classical signals to superconducting circuits \cite{lecocq2021control, youssefi2021cryogenic}, devices designed to transduce quantum states must satisfy a different and more stringent set of requirements \cite{zeuthen2020figures}. In spite of substantial recent progress, it remains an outstanding challenge to combine these transducers with superconducting qubits in a manner that is not destructive to the information stored in the qubit.  An impressive recent experiment demonstrated the transduction of photons from a superconducting qubit to the optical domain \cite{mirhosseini2020superconducting}, but the state of the qubit was destroyed by the optical pulses required for transduction.  Furthermore, such piezo-optomechanical transducers require a specific piezoelectric materials platform for qubit fabrication, complicating the introduction of some recent advances in modular circuit QED \cite{reagor2016quantum, campagne2020quantum, chakram2021seamless}.  

In this work, we use a mechanically-mediated electro-optic transducer to perform non-destructive optical readout of a transmon qubit embedded in a modular 3D circuit QED architecture.  The transducer imparts minimal backaction on the qubit, equivalent to only $\Delta n= (3\pm1)\times 10^{-3}$ photons in the microwave readout cavity dispersively coupled to the qubit. 
Although the transduction bandwidth is relatively narrow, the operation of the transducer is continuous rather than pulsed \cite{sahu2021quantum-enabled, mirhosseini2020superconducting, stockill2021ultra-low-noise}, enabling efficiency and repetition rates far exceeding the values demonstrated by a device that integrates the transducer and the qubit on the same chip \cite{mirhosseini2020superconducting}.  We then use the superconducting qubit as a non-Gaussian resource to characterise the quantum efficiency \cite{bultink2018general} with which we transduce signals from the circuit QED system to the optical domain. 
\begin{figure}
    \centering
    \includegraphics{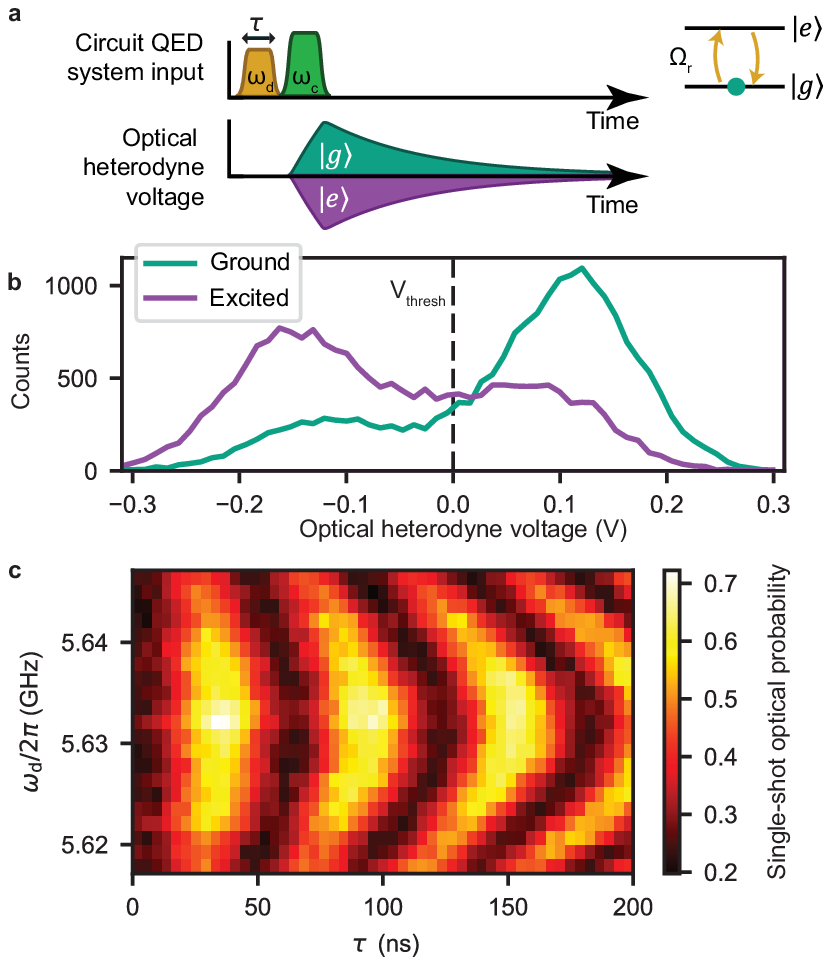}
    \caption{\textbf{Single-shot optical readout of a transmon qubit} (a) A microwave drive pulse with frequency $\omega_\textrm{d}$ near the qubit frequency and length $\tau$ is applied to drive Rabi oscillations in the qubit at rate $\Omega_\textrm{r}$.  A microwave readout pulse centred around $\omega_\textrm{c}$ is sent through the cavity and subsequently transduced to the optical domain for optical heterodyne detection.  The phase of the optical pulse depends on the state of the qubit. Pulse widths are not to scale for clarity.  (b) Histogram of optical heterodyne voltage when reading out the superconducting qubit through the electro-optic transducer.  A 15~\si{\mu\second} microwave pulse ($\sqrt{\bar{n}_\textrm{r}} = 19$~$\textrm{photons}^{1/2}$) is applied to the microwave cavity to optically read out the state of the superconducting qubit when preparing it in the ground state (teal curve) or excited state (purple curve), with $(\Gamma_\textrm{e}, \Gamma_\textrm{o})/2\pi = (0.5, 2.4)$~kHz.  The dashed line represents the voltage threshold $V_\textrm{thresh}$ for single-shot readout.   (c) Optically measured Rabi oscillations with the same readout pulse and transducer parameters as above.}  
    \label{fig:rabi}
\end{figure}

The experiment consists of two modular systems attached to the base plate of an optical-access dilution refrigerator at $T_\textrm{bp} \approx 40$~\si{\milli\kelvin}.  The first of these (Fig.~\ref{fig:fig1}(a)) is a 3D circuit QED system comprising a transmon qubit dispersively coupled to a quarter-wave coaxial cavity resonator \cite{reagor2016quantum}.  The microwave cavity has a total linewidth of $\kappa_\textrm{c}/2\pi = 380$~kHz, while the dispersive interaction between the qubit and the cavity causes a state-dependent shift of the cavity's resonant frequency by $2\chi/2\pi = 344$~kHz; thus $\kappa_\textrm{c}/2\chi \approx 1$, the optimal value for low-power dispersive readout of the qubit \cite{clerk2010introduction}.  The microwave cavity's resonant frequency $\omega_\textrm{c}$ can be tuned \textit{in situ} by a sapphire rod (see Fig.~\ref{fig:fig1}(a)) attached to a piezoelectric stepping module, enabling the microwave resonance of the circuit QED system to be brought into resonance with the transducer's microwave mode at $\omega_\textrm{e}/2\pi = 7.938$~GHz.

The other experimental module (Fig.~\ref{fig:fig1}(c)) is an electro-optic transducer containing microwave and optical resonators coupled to a single mode of a silicon nitride membrane with a mechanical resonance frequency of $\omega_\textrm{m}/2\pi = 1.45$~MHz \cite{higginbotham2018harnessing,ground}. Simultaneously applying strong microwave and optical pumps red-detuned by $\omega_\textrm{m}$ (see Fig.~\ref{fig:fig1}(e)) enhances the parametric coupling to the mechanical resonator \cite{higginbotham2018harnessing}. This enables microwave (optical) photons to be swapped through a beamsplitter interaction with phonons at a rate $\Gamma_\textrm{e}$ ($\Gamma_\textrm{o}$) far exceeding the intrinsic mechanical dissipation rate of $\gamma_\textrm{m} = 2\pi \times 0.11$~Hz. An incident signal on resonance with the microwave (optical) resonator will then be transduced to the optical (microwave) domain with the bandwidth of the process determined by the total damping rate $\Gamma_\textrm{T} = \Gamma_\textrm{e} + \Gamma_\textrm{o} + \gamma_\textrm{m}$.

Our optically mediated qubit readout scheme uses the dispersive interaction between the qubit and the microwave cavity, which causes the cavity's resonant frequency $\omega_\textrm{c} \pm \chi$ to depend on whether the qubit is prepared in the ground or excited state \cite{blais2004cavity}. Thus if a short microwave readout pulse centred at $\omega_\textrm{c}$ is applied to the microwave cavity input line (see Fig.~\ref{fig:fig1}(a)), the phase of the emitted microwave pulse will depend on the state of the qubit.  This microwave pulse is then routed to the electro-optic transducer through elements used to isolate the qubit from transducer backaction (see Fig.~\ref{fig:fig1}(b) and Methods), and the upconverted readout pulse at wavelength $\lambda=1084$~nm is demodulated using balanced heterodyne detection (Fig.~\ref{fig:fig1}(d)). During qubit readout experiments, the transducer pumps are applied continuously, while qubit preparation and readout pulses are repeated at intervals $T_\textrm{r}$ ranging from 0.4~ms to 2~ms depending on the bandwidth of the transducer. The bandwidth of the transducer filters the upconverted readout pulse as shown in the bottom row of  Fig.~\ref{fig:fig1}(f), which sets the upper limit on $T_\textrm{r}$.

To initialise the qubit state, a short qubit drive pulse with frequency $\omega_\textrm{d}$ and duration $\tau$ (gold pulse in Fig.~\ref{fig:rabi}(a)) can be applied to the input line of the circuit QED system to induce Rabi oscillations in the qubit at a rate $\Omega_\textrm{r}$. For $\tau = \pi/\Omega_\textrm{r}$, the qubit population is inverted, and prepared mainly in the excited state $|e\rangle$.  When no pulse is applied ($\tau=0$), the qubit is prepared mainly in the ground state $|g\rangle$.  State preparation is followed by a square readout pulse of length $T_\textrm{p} = 15$~\si{\micro\second} (green pulse in Fig.~\ref{fig:rabi}(a)).  The maximum useful length of the readout pulse, and hence the minimum pulse bandwidth, is determined by the lifetime of the qubit ($T_1 = 17$~\si{\micro\second}) \cite{gambetta2007protocols}.  The pulse then travels through the cavity and is upconverted as described above, and the demodulated optical signal is digitised and integrated to extract a single voltage encoding the state of the qubit.  

We first use this protocol to demonstrate single-shot optical readout of the superconducting qubit.  By recording multiple voltage traces to form histograms of the qubit state-dependent optical heterodyne voltage, we can estimate the single-shot probability $P(e)$, and choose a voltage threshold $V_\textrm{thresh}$ (see dashed line in Fig.~\ref{fig:rabi}(b)) to maximise the fidelity $F_\textrm{opt} = 1-P(e|g) - P(g|e)$, where $P(e|g)$ ($P(g|e)$) is the probability of measuring $|e\rangle$ ($|g\rangle$) given that $|g\rangle$ ($|e\rangle$) was prepared. In Fig.~\ref{fig:rabi}(b) we show histograms of the optical heterodyne voltage when preparing the qubit in either $|e\rangle$ or $|g\rangle$.  The amplitude of the microwave readout pulse incident on the circuit QED system's microwave cavity is $\sqrt{\bar{n}_\textrm{r}} = 19$~$\textrm{photons}^{1/2}$, and the electro-optic transducer is operated continuously with $(\Gamma_\textrm{e}, \Gamma_\textrm{o})/2\pi = (0.5, 2.4)$~kHz to transduce the emitted microwave field to the optical domain for detection. A bimodal distribution is clearly visible in each histogram, with the two modes corresponding to the ground and excited states of the qubit.  We find a maximum optical readout fidelity of $F\approx 0.4$, limited by inefficient measurement and residual excited state population in the qubit (see Supplementary Information).    

We can then use single-shot optical readout to measure Rabi oscillations of the superconducting qubit and demonstrate the stability of the optical readout.  To measure Rabi oscillations we vary the qubit drive pulse length $\tau$ and the pulse frequency $\omega_\textrm{d}$ (Fig.~\ref{fig:rabi}(c), with the same pulse amplitude and transducer damping rates as in Fig.~\ref{fig:rabi}(b)).  This measurement was taken over approximately 1.5 hours using a single threshold value chosen at the beginning of the experiment, indicating stability of the optical cavity lock and electro-optic transducer over this period of time.  
\begin{figure}
    \centering
    \includegraphics{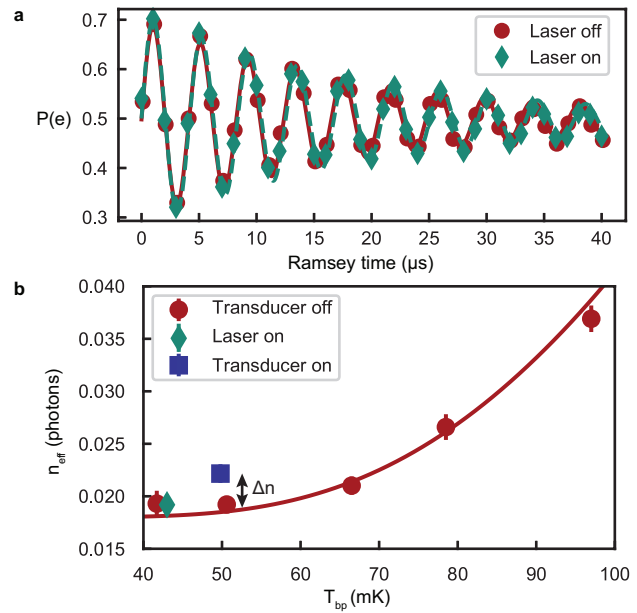}
    \caption{\textbf{Transducer backaction on the qubit} (a) Microwave-domain Ramsey experiment with the transducer off (($\Gamma_\mathrm{o},\Gamma_\mathrm{e}) = (0,0)$; red circles) and with the laser on (($\Gamma_\mathrm{o},\Gamma_\mathrm{e})/2\pi = (0,5.0)$~kHz; cyan diamonds).  The lines are fits to the data.  (b) Effective thermal occupancy of the circuit QED system's microwave cavity as a function of the dilution refrigerator base plate temperature $T_\textrm{bp}$.  Red circles are data obtained with the transducer off, while the red line is theory.
    The cyan diamond indicates the occupancy of the microwave cavity with $\Gamma_\textrm{o}/2\pi = 5.0$~kHz, indicating no additional dephasing of the qubit from the optical field.  The blue square is the effective occupancy while operating the transducer at $(\Gamma_\textrm{e}, \Gamma_\textrm{o})/2\pi = (1.1, 5.0)$~kHz.  There is a small amount of excess backaction equivalent to $\Delta n = (3\pm1) \times 10^{-3}$ photons in the microwave cavity. All error bars represent one standard deviation.} 
    \label{fig:backaction}
\end{figure}

Next, we characterise the effects of backaction from the transducer on the qubit. An ideal electro-optic transducer performs unitary operations on the microwave and optical fields, and need not add noise or impart backaction during the transduction process \cite{caves1982quantum}.  
However, under real experimental conditions the effective thermal bath $n_\textrm{eff}$ coupled to the circuit QED system imparts backaction on the qubit in the form of shot noise from microwave photons randomly arriving in the cavity.  For $n_\textrm{eff}\ll1$ these photons cause dephasing of the qubit of the form \cite{clerk2007using}
\begin{equation}
    \label{eq:dephasing}
    \Gamma_\phi = \frac{\kappa_\textrm{c}\chi^2}{\frac{\kappa_\textrm{c}^2}{4} + \chi^2}n_\textrm{eff},
\end{equation}
where we can split the effective occupancy $n_\textrm{eff} = n_\textrm{th} + \Delta n$ into a thermal component $n_\textrm{th}$ and a component describing excess backaction $\Delta n$ from the transducer.  It is desirable to minimise this excess backaction as it can limit the fidelity of error correction and post-selection operations on the qubit \cite{hatridge2013quantum}.  We are able to eliminate most of the backaction from the transducer using a combination of circulators and interferometric cancellation \cite{mallet2011quantum} of the strong microwave pump required for efficient transduction (see Fig.~\ref{fig:fig1}(b) and Methods).

We first measure the effect of the optical pump on the circuit QED system, with the microwave pump off.  In Fig.~\ref{fig:backaction}(a) a microwave-domain Ramsey experiment is performed to measure the coherence time $T_2$ of the qubit when the laser is off (red circles) and when it is on with $\Gamma_\textrm{o}/2\pi = 5.0$~kHz (cyan diamonds), achieved with $11$~mW of circulating power in the optical cavity.  Heating of the circuit QED system by the laser would dephase the qubit and reduce its coherence time.  We find $T_\textrm{2,on} = 20.7\pm 0.2$~\si{\micro\second} and $T_\textrm{2,off} = 20.4\pm 0.2$~\si{\micro\second}, showing no measurable change in the qubit coherence time due to laser illumination.  

We can expand the scope of this measurement, and additionally measure the lifetime of the qubit to determine its dephasing rate $\Gamma_\phi = T_\phi^{-1} = T_2^{-1} - \frac{1}{2}T_1^{-1}$, and infer the effective occupancy $n_\textrm{eff}$ of the circuit QED system's microwave cavity using Eq.~\eqref{eq:dephasing}.  In Fig.~\ref{fig:backaction}(b) we plot this effective occupancy as the temperature of the dilution refrigerator's base plate $T_\textrm{bp}$ is varied.  At low temperature, the inferred thermal occupancy plateaus at $n_\textrm{th} \approx 0.019$ photons ($T_\textrm{eff} = 95$~mK), though this plateau may be caused by intrinsic sources of qubit dephasing other than an elevated temperature \cite{krantz2019quantum}.  The cyan diamond in Fig.~\ref{fig:backaction}(b) demonstrates that there is no measurable excess backaction from the laser ($\Gamma_\textrm{o}/2\pi = 5.0$~kHz) on the qubit.  We then turn on the microwave pump ($\Gamma_\textrm{e}/2\pi = 1.1$~kHz, leaving $\Gamma_\textrm{o}$ unchanged) and the blue square in Fig.~\ref{fig:backaction}(b) indicates excess backaction of $\Delta n = (3\pm1)\times 10^{-3}$ photons.  This excess backaction comes largely from local heating of microwave components by the strong microwave pump with incident power $P_\textrm{inc} = -45$~dBm---see Methods for details.  We emphasise that these backaction measurements are taken at the largest values of $\Gamma_\textrm{o}$ and $\Gamma_\textrm{e}$ used in this work, and backaction will likely be even lower when operating at smaller $\Gamma_\textrm{e}$.

\begin{figure}
    %\centering
    \includegraphics{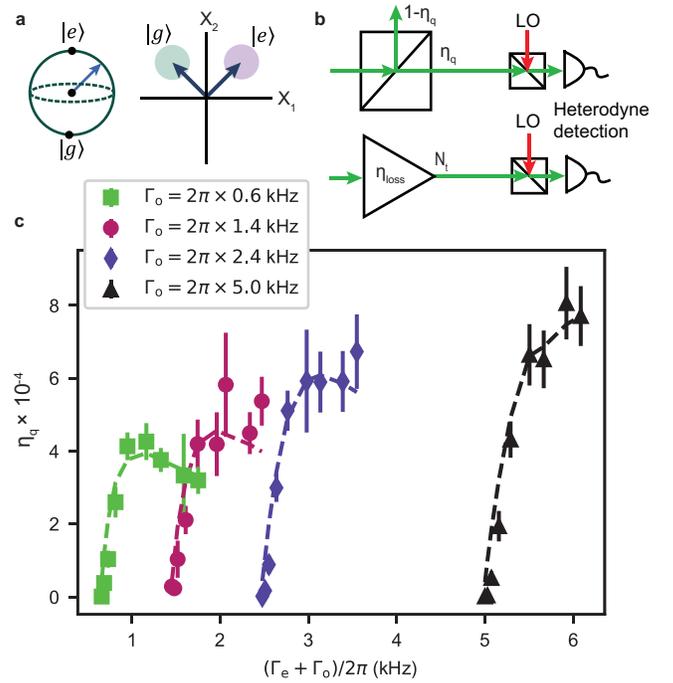}
    \caption{\textbf{Quantum efficiency of the optically mediated qubit readout} (a)  During dispersive readout, the state of the qubit is mapped onto one of two (microwave) coherent states, which is then upconverted and detected in heterodyne.  The quantum efficiency $\eta_\textrm{q}$ is a measure of how well the optical measurement apparatus can distinguish these two coherent states.  (b)   The optical measurement apparatus can be modelled as a beamsplitter with effective transmissivity $\eta_\textrm{q}$ followed by an ideal optical heterodyne detector.  Alternatively, it can be modelled as an amplifier with gain $\eta_\textrm{loss}=\eta_\textrm{q}/\eta_\textrm{noise}$ and output-referred added noise $N_\textrm{t}$, whose output is directed towards an ideal optical heterodyne detector. (c) The quantum efficiency is calibrated by comparing the signal-to-noise ratio of an optically mediated qubit readout measurement to the total measurement-induced dephasing.  For  $20~\textrm{Hz}\leq\Gamma_\mathrm{e}\leq1100~\textrm{Hz}$ we vary $\Gamma_\textrm{e}$ for several fixed values of $\Gamma_\textrm{o}$ and perform qubit readout measurements.  The points are data from these measurements, while the dashed lines are a model using independently measured parameters of the electro-optic transducer.  The model includes all sources of inefficiency such as loss, added noise, and finite transduction bandwidth. All error bars represent one standard deviation.}
    \label{fig:fig4}
\end{figure}

The final figure of merit for optically mediated qubit readout is the quantum efficiency $\eta_\textrm{q}$, which is a measure of how well the optical measurement apparatus can distinguish the coherent states encoding the state of the qubit, and can be characterised $\textit{in situ}$ by using the superconducting qubit as a non-Gaussian resource \cite{bultink2018general, rosenthal2021efficient, touzard2019gated}.  As measurement photons transit the cavity and extract information about the state of the qubit, the phase coherence of the qubit is necessarily destroyed \cite{gambetta2006qubit}.  This establishes a fundamental relationship between the signal-to-noise ratio of a dispersive measurement (see Fig.~\ref{fig:fig4}(a)) and the measurement-induced dephasing due to the qubit readout pulse \cite{bultink2018general}---see Methods. An imperfect readout apparatus degrades the signal-to-noise ratio further, and thus reduction of the signal-to-noise ratio below this fundamental bound can be used to measure the loss and added noise of the readout process. We model the measurement apparatus as a beamsplitter with effective transmissivity $\eta_\textrm{q}$ (see upper row in Fig.~\ref{fig:fig4}(b)), with an ideal optical heterodyne detector at the beamsplitter output, combining the effects of loss and added noise into a single performance metric (see Methods for details).  

In Fig.~\ref{fig:fig4}(c) we measure the quantum efficiency while varying the electromechanical damping $\Gamma_\textrm{e}$ for several different fixed values of the optomechanical damping rate $\Gamma_\textrm{o}$.   For $\Gamma_\textrm{e}/\Gamma_\textrm{o}\ll1$ the quantum efficiency decreases due to mismatched damping rates \cite{ground}, while the plateau in efficiency is primarily due to noise emitted by the LC circuit and $\Gamma_\textrm{e}$-dependent LC circuit loss. From these measurements we obtain a maximum quantum efficiency of $\eta_\textrm{q} \approx 8 \times10^{-4}$.  The dashed lines in Fig.~\ref{fig:fig4}(c) are obtained from a state-space model using measured transducer parameters---see Supplementary Information. 

To more directly compare our optically mediated qubit readout to existing results, we can instead describe the optical measurement apparatus as an effective amplifier with gain $\eta_\textrm{loss} = \eta_\textrm{q}/\eta_\textrm{noise}$ and two-quadrature added noise $N_\textrm{t} = \eta_\textrm{noise}^{-1}-1$ at its output (see Methods and lower row in Fig.~\ref{fig:fig4}(b)), followed by an ideal optical heterodyne detector.
Thus the total noise at the input of the ideal optical heterodyne detector $N_\textrm{det} = 1+N_\textrm{t}$ consists of the sum of vacuum fluctuations, the added noise of this ideal heterodyne detector, and the transducer added noise $N_\textrm{t}$.

We obtain a maximum transmission efficiency of $\eta_\textrm{loss} = 1.9\times 10^{-3}$, which captures \textit{all} sources of loss between the circuit QED system and the optical detector, and is more than two orders of magnitude larger than the best demonstrated efficiency for transduction of states from a superconducting qubit to the optical domain \cite{mirhosseini2020superconducting}.  This maximum value of $\eta_\textrm{loss}$ is the product of many individual contributions, and includes the transducer efficiency as defined in Ref.~\cite{higginbotham2018harnessing} ($\eta_\textrm{t} = 0.19$), optical detection efficiency ($\eta_\textrm{opt} = 0.28$), microwave transmission loss ($\eta_\textrm{mic} = 0.34$), and inefficiency due to the bandwidth of the readout pulse exceeding that of the transducer ($\eta_\textrm{bw} = 0.15$)---see Methods.  The noise at the input of the optical heterodyne detector is relatively high at this operating point ($N_\textrm{det}=1+N_\textrm{t} =2.4$~photons) owing primarily to technical noise emitted by the LC circuit, which is caused by the application of the strong microwave pump and is present even in the absence of laser light \cite{ground}.  It is this added noise that is the major factor preventing quantum-enabled transduction \cite{zeuthen2020figures}---see Methods.

However, with only moderate improvements in the bandwidth and added noise of the electro-optic transducer, the upconversion of non-trivial quantum states from a superconducting qubit is well within reach.  Using standard techniques \cite{reagor2016quantum, place2021new}, the lifetime of the qubit in our circuit QED system can straightforwardly be improved to $T_1>100$~\si{\micro\second}.  Moreover, the vacuum electromechanical coupling $g_\textrm{e}$ was smaller in this device than in previously measured transducers \cite{ground}, suggesting that the transduction bandwidth can easily be improved to $\sim10$~kHz by enhancing this coupling.  The combination of spectrally narrower pulses emerging from the circuit QED system and increased $g_\textrm{e}$ would largely remove bandwidth as a limiting factor, and enhancements in $g_\textrm{e}$ would also reduce microwave noise by reducing the  pump power required to achieve a given damping \cite{ground}.  With $\Gamma_\textrm{o} = \Gamma_\textrm{e} = 2\pi\times5$~kHz, $T_1 = 100$~\si{\micro\second}, and $N_\textrm{t} \ll 1$ we expect a quantum efficiency of $\eta_\textrm{q} \approx 0.1$ and the ability to transmit heralded quantum states from superconducting qubits exhibiting non-classical statistics. 
Two such systems would then enable remote entanglement of superconducting qubits over kilometer-scale distances. 
%TC:ignore
%

\section{Methods}

\section{Circuit QED system} The circuit QED system consists of a planar superconducting transmon qubit coupled to the fundamental mode of a seamless 3D aluminium cavity.  The qubit was fabricated on a $14 \times 2~\text{mm}^2$ C-plane sapphire chip. Using a single e-beam lithography step, the Dolan bridge process \cite{dolan1977offset} was used to fabricate an \ch{Al-AlO_{x}-Al} Josephson junction with area of~\SI[product-units=power]{150 x 250}{\nm}.  The qubit capacitance comes from two rectangular $500 \times 700$ \si{\micro\meter} pads separated by $200$~\si{\micro\meter}.    

The cavity, a $40$~mm long circular waveguide with radius $4.7$~mm, was machined out of a single piece of $99.999$\% purity aluminium. The waveguide is open at one end, with a $7.25$~mm long post with 1.5~mm radius at the other end; this geometry defines the fundamental mode as a quarter-wave resonance with evanescent coupling to the waveguide above \cite{reagor2016quantum}.  To reduce surface loss, the cavity was etched in Transene aluminium etchant Type A for 24 hours at ambient temperature. A sapphire rod with radius $2$~mm is epoxied to a copper plate and fastened to a piezoelectric stepping module above the waveguide with a maximum travel of 5~mm.  This assembly allows the sapphire rod to be translated along the centre of the waveguide to tune the fundamental mode's resonant frequency $\omega_\textrm{c}$ \textit{in situ} by up to $1.7$~GHz.  The cavity mode has a total linewidth of $\kappa_\textrm{c} = \kappa_\textrm{c,w} + \kappa_\textrm{c,int} + \kappa_\textrm{c,ext} = 2\pi \times 380$~kHz and is strongly overcoupled to the transmission line connected to the electro-optic transducer.

The circuit QED system can be described by the Jaynes-Cummings model with interaction Hamiltonian $\hat{H}_\textrm{int}^\textrm{JC} = \hbar g_\textrm{qc}(\hat{d}^\dagger\hat{\sigma}^-  +\hat{d}\hat{\sigma}^+)$.  Here, the resonator has creation (annihilation) operator $\hat{d}^\dagger$ $(\hat{d})$, and can swap excitations with the qubit via the raising $(+)$ and lowering $(-)$ operators $\hat{\sigma}_\pm$.  When the qubit is far detuned from the resonator ($g_\textrm{qc} \ll |\Delta_\textrm{qc}|$, where $\Delta_\textrm{qc} = \omega_\textrm{q} - \omega_\textrm{c}$), the Jaynes-Cummings model can be described under the dispersive approximation $\hat{H}_\textrm{int}^\textrm{JC} = -\hbar\chi\hat{\sigma}_\textrm{z}\hat{d}^\dagger \hat{d}$. The dispersive shift $\chi$ is given by $\chi = g_\textrm{qc}^2\nu/\left(\Delta_\textrm{qc}(\Delta_\textrm{qc}-\nu)\right)$ \cite{koch2007charge}, where the qubit anharmonicity $\nu$ encodes the contribution from higher energy levels of the transmon qubit.  See Extended Data Table~\ref{tab:transducer} for a list of parameters describing the circuit QED system.  

The qubit lifetime ($T_1 = 17$~\si{\micro\second}) is far from Purcell-limited \cite{houck2008controlling}, likely due to participation of lossy dielectrics that remain after e-beam lithography and aluminium deposition.  The coherence time of the qubit is $T_2^\textrm{R} \approx T_2^\textrm{echo} \approx 20$~\si{\micro\second}, and no difference is seen when it is measured using a Hahn echo versus a standard Ramsey sequence.   

The circuit QED system is contained in a magnetic shield comprising an outer shield of high-permeability Amumetal 4K and an inner shield made of pure aluminium in order to minimise stray magnetic fields near the circuit QED system during cooling of the device through its critical temperature.

\section{Electro-optic transducer}  

The electro-optic transducer consists of a microwave resonator and an optical resonator coupled to a single mode of a micromechanical oscillator.  It is described by the Hamiltonian
\begin{equation}
    \frac{\hat{H}}{\hbar} = \omega_\textrm{o} \hat{a}^\dagger \hat{a} + \omega_\textrm{e}\hat{b}^\dagger \hat{b} + \omega_\textrm{m}\hat{c}^\dagger \hat{c} + (g_\textrm{o}\hat{a}^\dagger \hat{a} + g_\textrm{e}\hat{b}^\dagger \hat{b})(\hat{c}+\hat{c}^\dagger),
\end{equation}
where $\hat{a}$, $\hat{b}$ and $\hat{c}$ are the annihilation operators for the optical, microwave and mechanical modes, respectively, and  $g_\textrm{o}$ $(g_\textrm{e})$ is the vacuum optomechanical (electromechanical) coupling rate.  

The mechanical oscillator is a 100~nm thick, 500~\si{\mu\meter} wide square silicon nitride membrane suspended from a silicon chip. Phononic shielding is patterned into the silicon chip, isolating the membrane mode used for transduction, with resonant frequency $\omega_\textrm{m} = 2\pi \times 1.45$~MHz, from vibrational modes of the chip. This shielding results in a quality factor $Q_\textrm{m}=1.3\times10^7$ for the mode of interest, or equivalently an energy dissipation rate $\gamma_\textrm{m} = 2\pi \times 0.11 $~Hz. 

The microwave resonator is a superconducting flip-chip LC circuit whose capacitance is modulated by the motion of a superconducting pad deposited on the membrane. The vacuum electromechanical coupling is determined by the capacitor gap spacing between this pad and a second nearby chip hosting the rest of the circuit, which is $\sim300$~nm for most devices. In the transducer used in this work the capacitor gap was unusually large, resulting in a low vacuum electromechanical coupling rate $g_\textrm{e} = 2\pi \times 1.6$~Hz. The LC circuit is coupled to a microwave transmission line at rate $\kappa_\textrm{e,ext} = 2\pi \times 1.42$~MHz, and its total linewidth $\kappa_\textrm{e}$ varies between 1.6~MHz and 2.7~MHz due to dependence of the internal loss $\kappa_\textrm{e,int}$ on the power of the microwave pump used to mediate transduction.  

The optical resonator is a Fabry-P\'erot cavity defined by high-reflectivity ion beam sputtered mirror coatings deposited by FiveNine Optics, which are chosen to be very asymmetric such that the cavity mode couples preferentially out the front mirror with external coupling  $\kappa_\textrm{o,ext}= 2\pi\times2.12$~MHz. The total cavity linewidth is $\kappa_\textrm{o}=\kappa_\textrm{o,ext}+\kappa_\textrm{o,int}=2\pi \times 2.68$~MHz, where $\kappa_\textrm{o,int}$ includes scattering and absorption losses as well as transmission through the back mirror. The membrane is positioned at a maximum in the intensity gradient of the intracavity light, yielding a vacuum coupling rate $g_\textrm{o} = 2\pi \times 60$~Hz. The heterodyne mode matching factor quantifying the overlap of the propagating optical cavity mode and the local oscillator beam (LO) is $\epsilon = 0.80$.  

During operation, strong microwave and optical pumps are applied near the corresponding electromagnetic resonances to enhance the vacuum coupling rates $g_\textrm{o}$ and $g_\textrm{e}$ \cite{andrews2014bidirectional}.  Transforming to a rotating frame to remove the free evolution of the fields, $\hat{a}\rightarrow \hat{a}e^{-i\omega_\textrm{o}t}$, $\hat{b}\rightarrow \hat{b}e^{-i\omega_\textrm{e}t}$ and $\hat{c}\rightarrow \hat{c}e^{-i\omega_\textrm{m}t}$, the Hamiltonian can then be linearised around the strong pumps to yield
\begin{equation}
    \frac{\hat{H}_\textrm{int}^\textrm{lin}}{\hbar} = g_\textrm{o}\bar{a}(\hat{a}^\dagger + \hat{a})(\hat{c}+ \hat{c}^\dagger)+g_\textrm{e}\bar{b}(\hat{b}^\dagger + \hat{b})(\hat{c}+\hat{c}^\dagger),
    \label{eq:Hint}
\end{equation}
where $\bar{a}$ ($\bar{b}$) is the optical (microwave) mode coherent state amplitude due to the incident pump.  Both pumps are red-detuned from the respective resonances by $\omega_\textrm{m}$, resonantly enhancing the optomechanical and electromechanical beamsplitter terms in the Hamiltonian \cite{aspelmeyer2014cavity}. In the resolved sideband limit at this optimal detuning, the electromechanical (optomechanical) damping rate is then given by $\Gamma_\textrm{o} = 4g_\textrm{o}^2\bar{a}^2/\kappa_\textrm{o}$ $\left(\Gamma_\textrm{e} = 4g_\textrm{e}^2\bar{b}^2/\kappa_\textrm{e}\right)$. The transducer bandwidth is given by $\Gamma_\textrm{T} = \Gamma_\textrm{e} + \Gamma_\textrm{o} + \gamma_\textrm{m}$.

The electro-optic transducer parameters are summarised in Extended Data Table~\ref{tab:transducer}. See Ref.~\cite{ground} for a more detailed description of the fabrication, assembly, and characterisation of the electro-optic transducer, as well as the theory of transducer operation.

\section{Superconducting qubit readout signal-to-noise ratio}  %Dispersive readout has previously been used to characterise the quantum efficiency of parametric amplifiers used in superconducting qubit experiments \cite{bultink2018general, rosenthal2021efficient,touzard2019gated}.  
In a dispersive measurement, a coherent state $|\alpha\rangle$ is entangled with the state of the qubit.  If initially the qubit is prepared in a superposition state $|\psi_\textrm{q}\rangle = \frac{1}{\sqrt{2}}(|e\rangle + |g\rangle)$, its interaction with the driven cavity mode causes it to evolve into the state
\begin{equation}
|\psi_\textrm{m}\rangle = \frac{1}{\sqrt{2}}\left(|e\rangle |\alpha_\textrm{e}\rangle + |g\rangle |\alpha_\textrm{g} \rangle\right),
\end{equation}
where $|\alpha_\textrm{e}\rangle$ and $|\alpha_\textrm{g}\rangle$ are coherent states with equal magnitude but phases shifted by $\theta_\pm = \pm\arctan{2\chi/\kappa_\textrm{c}}$, with the $+$ $(-)$ sign corresponding to the qubit in $|g\rangle$ ($|e\rangle$). The readout amplitude $\sqrt{\bar{n}_\textrm{r}} = |\alpha|\sin \theta $ determines the separation in phase space between these two coherent states.  The degree to which we can resolve the two Gaussian distributions in phase space is quantified by the signal-to-noise ratio (SNR): loss degrades the SNR by reducing the phase space separation, while noise degrades the SNR by increasing the variance of each distribution.  We define the $\textrm{SNR}$ as
\begin{equation}
    \textrm{SNR} = \frac{|\mu_\textrm{e} - \mu_\textrm{g}|}{\sigma_\textrm{ge}},
\end{equation}
where $\sigma_\textrm{ge} = \sigma_{e} = \sigma_{g}$ is the standard deviation of the Gaussian-distributed optical heterodyne voltage corresponding to the qubit in either the ground or the excited state, and $\mu_\textrm{g}$ and $\mu_\textrm{e}$ are the mean values of the two distributions.  Noise $N_\textrm{t}$ gets added by the transducer, and some of the signal is lost such that $|\alpha| \rightarrow \eta_\textrm{loss} |\alpha|$; thus the SNR can be written as
\begin{equation}
    \label{eq:SNR2}
    \textrm{SNR} = \frac{2\sqrt{2}\eta_\textrm{loss}|\alpha|\sin \theta}{\sqrt{1+N_\textrm{t}}}.  
\end{equation}
\section{Measurement-induced dephasing and the quantum efficiency}
Reading out the state of the qubit through the cavity necessarily causes fluctuations in the frequency of the qubit due to the dispersive interaction between the qubit and the cavity $H_\textrm{int} = -\chi  \hat{\sigma}_\textrm{z}\hat{d}^\dagger \hat{d}$.  This leads to measurement-induced dephasing \cite{gambetta2006qubit}, destroying the coherence of any initial superposition. The off-diagonal elements of the qubit's density matrix after this interaction are given by $\rho_\textrm{ge} = \rho_\textrm{ge}(0)e^{-2|\alpha|^2 \sin^2(\theta)} = e^{-2\bar{n}_\textrm{r}}$ \cite{rosenthal2021efficient}. 

To demonstrate measurement-induced dephasing we inject a weak measurement pulse into a Ramsey sequence (see Extended Data Fig.~\ref{fig:quant_eff_ext}(a)), and measure the resulting amplitude of the Ramsey oscillations---where $\rho_\textrm{ge}=\langle\hat{\sigma}_\textrm{z}\rangle/4$---using a strong projective measurement of the qubit.  Extended Data Fig.~\ref{fig:quant_eff_ext}(b) shows the results of these measurements, using the microwave readout chain and varying the strength of the weak $15$~\si{\micro\second} readout pulse. We observe a clear Gaussian decay of $\rho_\textrm{ge}$ as a function of readout amplitude $\sqrt{n_\textrm{r}}$. 

This measurement-induced dephasing quantifies the amplitude of the readout pulse inside of the circuit QED system's microwave cavity, while the SNR quantifies the distinguishability of the two (thermal) coherent states at the optical detector.  Thus these two quantities are fundamentally related to each other via the quantum efficiency $\eta_\textrm{q} = \frac{\eta_\textrm{loss}}{1+N_\textrm{t}}$ \cite{bultink2018general}, which describes the optical measurement apparatus as an effective beamsplitter between the circuit QED system and an ideal optical heterodyne detector.

\section{Technical details of the quantum efficiency calibration}
To control the readout amplitude $\sqrt{n_\textrm{r}}$ a drive voltage $V$ is varied via an IQ mixer.  The SNR and Gaussian decay of $\rho_\textrm{ge}$ can be recast in terms of this drive voltage as: $\textrm{SNR} = aV$ and $\rho_\textrm{ge}=\rho_\textrm{ge}(0)e^{-\frac{V^2}{2\sigma^2}}$.  As shown in Ref.~\cite{bultink2018general}, the quantum efficiency is then a function of the slope of the SNR (see Extended Data Fig.~\ref{fig:quant_eff_ext}(c)) and the width of the Gaussian distribution $\sigma$:
\begin{equation}
    \eta_\textrm{q} = \frac{\sigma^2 a^2}{2}, 
\end{equation}
where in this convention an ideal heterodyne detector would have $\eta_\textrm{q} = 1$
\section{Contributions to the quantum efficiency}  The quantum efficiency of the optically mediated qubit readout apparatus can be split into seven main components, $\eta_\textrm{q} = \eta_\textrm{bw} \eta_\textrm{t} \eta_\textrm{G}  \eta_\textrm{mic} \eta_\textrm{opt}\eta_\textrm{cav}\eta_\textrm{noise}$. We describe each contribution below, and the contributions are also summarised in Extended Data Table~\ref{tab:quantum_eff}.

\textbf{Bandwidth limitations.}
A significant contribution to $\eta_\textrm{q}$ is due to the mismatch in bandwidth between the transducer and the readout pulse emitted by the qubit cavity.  In the rotating wave approximation, and approximating the circuit QED system's output as a square pulse with width $T_\textrm{p}$, we obtain
\begin{equation}
    \eta_\textrm{bw} \approx 1 - 2\frac{\left(1 - e^{-\Gamma_\textrm{T}T_\textrm{p}/2}\right)}{\Gamma_\textrm{T}T_\textrm{p}}.
    \label{eq:eta_bw}
\end{equation} 
Solving for the full dynamics without the above approximations gives a contribution to the efficiency in the range $0.02 < \eta_\textrm{bw} < 0.15$ over the range of $\Gamma_\textrm{T}$ values plotted in Fig.~\ref{fig:fig4}(c) in the main text.

\textbf{Transducer efficiency.} The relatively low electromechanical coupling in this device ($g_\textrm{e} = 2\pi \times 1.6$~Hz) limits the maximum achievable electromechanical damping rate to $\Gamma_\textrm{e} = 2\pi\times1.1$~kHz.  To maximise bandwidth we often operate the transducer in a mismatched mode with $\Gamma_\textrm{o}\gg\Gamma_\textrm{e}$, but this comes at a cost to the narrowband signal efficiency of the transducer \cite{ground}:
\begin{equation}
    \eta_\textrm{t} = \epsilon \frac{\kappa_\textrm{o,ext}}{\kappa_\textrm{o}}\frac{\kappa_\textrm{e,ext}}{\kappa_\textrm{e}}\frac{4\Gamma_\textrm{e}\Gamma_\textrm{o}}{\Gamma_\textrm{T}^2},
    \label{eq:eta_conv}
\end{equation}
which sets an upper bound on the efficiency with which broadband signals can be transduced. Eq.~\eqref{eq:eta_conv} also depends on $\Gamma_\textrm{e}$ implicitly due to the pump power-dependent LC circuit loss \cite{ground}. This contribution is responsible for the sharp drop in efficiency for $\Gamma_\textrm{e}\ll\Gamma_\textrm{o}$ on each curve in Fig.~\ref{fig:fig4}(c) in the main text, and contributes to the plateau in $\eta_\textrm{q}$ at high $\Gamma_\textrm{e}$ for each fixed value of $\Gamma_\textrm{o}$.

\textbf{Transducer gain} Due to the finite sideband resolution of the electro-optomechanical system \cite{aspelmeyer2014cavity}, the transducer has gain \cite{andrews2014bidirectional}
\begin{equation}
\eta_\textrm{G}=\left(1+\left(\frac{\kappa_\textrm{e}}{4\omega_\textrm{m}}\right)^2\right)\left(1+\left(\frac{\kappa_\textrm{o}}{4\omega_\textrm{m}}\right)^2\right).
\label{eq:transd_gain}
\end{equation}  
This two-quadrature gain is undesirable for transduction, as it is necessarily accompanied by unwanted added noise \cite{caves1982quantum}, but it is a small effect, varying in the range $1.3 < \eta_\textrm{G} < 1.5$ with microwave pump power.  

\textbf{Microwave transmission loss between the circuit QED system and the transducer.}
Our system also permits qubit readout through a microwave readout apparatus, shown in Fig.~\ref{fig:fig1}(b) of the main text. Its quantum efficiency $\eta_\textrm{q}^\textrm{mic}$ is defined to include all sources of loss and noise that affect the readout pulse as it emerges from the circuit QED system's cavity, reflects off the transducer's LC circuit, and is detected using the microwave heterodyne measurement chain. Measuring $\eta_\textrm{q}^\textrm{mic}$ allows us to back out the microwave transmission loss $\eta_\textrm{mic}$ between the circuit QED system and the transducer, which also contributes to $\eta_\textrm{q}$.  As shown in Extended Data Fig.~\ref{fig:mic_loss}, $\eta_\textrm{q}^\textrm{mic}$ decreases significantly with increasing $\Gamma_\textrm{e}$---mainly due to $\Gamma_\textrm{e}$-dependent reflection loss.  This data is fit to a model (purple curve in Extended Data Fig.~\ref{fig:mic_loss}) that includes the effects of $\Gamma_\textrm{e}$-dependent LC circuit reflection loss, noise emitted from the microwave port of the transducer, and independently calibrated microwave measurement chain added noise referred to the transducer output \cite{ground}, with $\eta_\textrm{mic}$ as the only free parameter, yielding $\eta_\textrm{mic} = 0.34$ (4.7 dB).  This relatively high loss can be mitigated by reducing the total number of microwave connectors in the signal path, switching to superconducting cables, and removing filters from the signal path and placing them instead on the pump and cancellation lines.

Using the quantum efficiency of the optical measurement apparatus and a state-space model (see Supplementary Information) we perform a fit to $\eta_\textrm{q}$ as $\Gamma_\textrm{e}+\Gamma_\textrm{o}$ is varied, with $\eta_\textrm{mic}$ as the only free parameter.  This yields an \textit{apparent} microwave loss of $\eta_\textrm{mic}^\textrm{app}= 0.17$, in slight tension with $\eta_\textrm{mic}=0.34$ inferred from the microwave measurement described above.  This discrepancy may be due to drifts in the optical cavity mode matching or microwave chain gain over a several month time scale.  

\textbf{Optical detection efficiency.}
The optical detection efficiency is given by the product of factors encoding optical transmission losses between the transducer and the balanced heterodyne detector, the inefficiency of the heterodyne detector itself, and the detector's dark noise: together these three factors yield $\eta_\textrm{opt} = 0.28$. This figure excludes optical cavity losses and imperfect mode matching, which are included in the transducer efficiency, Eq.~\eqref{eq:eta_conv}.

\textbf{Circuit QED system cavity loss.} Through the measured attenuation on the lines, we bound the sum of the of the weak port coupling and the internal loss to $\kappa_\textrm{c,int} + \kappa_\textrm{c,w}<2\pi\times15$~kHz.  Thus $\eta_\textrm{cav} = 1-\frac{\kappa_\textrm{c,int} + \kappa_\textrm{c,w}}{\kappa_\textrm{c}}>0.96$ does not significantly affect $\eta_\textrm{q}$.

\textbf{Added noise.}
Here we show how the noise measured at the input of the optical heterodyne detector, or equivalently at the output of an effective amplifier with gain $\eta_\textrm{loss}$, can alternatively be expressed as a contribution to the quantum efficiency $\eta_\textrm{q}$ of the optically mediated readout apparatus, modelled as an effective beamsplitter (see Fig.~\ref{fig:fig4}(b) in the main text).  

The voltage records obtained from optical heterodyne detection of the upconverted qubit readout pulse can be written in the form
\begin{equation}
I_{|k\rangle}(t) = \sqrt{G_\textrm{o}}\left(\sqrt{\kappa_\textrm{c}\eta_\textrm{loss}}\textrm{Re}(\alpha_{|k\rangle}(t)) + \hat{\zeta}_\textrm{I}(t)\right)
\label{eq:Iamp}
\end{equation}
\begin{equation}
Q_{|k\rangle}(t) = \sqrt{G_\textrm{o}}\left(\sqrt{\kappa_\textrm{c}\eta_\textrm{loss}}\textrm{Im}(\alpha_{|k\rangle}(t)) + \hat{\zeta}_\textrm{Q}(t)\right),
\label{eq:Qamp}
\end{equation}
where $\sqrt{G_\textrm{o}}$ is an overall gain factor and the index $k = \{g,e\}$ labels whether the qubit was prepared in the ground state or the excited state.  $\eta_\textrm{loss} =\eta_\textrm{bw} \eta_\textrm{t} \eta_\textrm{G}  \eta_\textrm{mic} \eta_\textrm{opt}\eta_\textrm{cav}$ includes \textit{all} sources of loss between the circuit QED system and the ideal optical detector, while $\hat{\zeta}_\textrm{I}$ and $\hat{\zeta}_\textrm{Q}$ are noise operators whose autocorrelation functions include contributions from vacuum fluctuations, the added noise of an ideal heterodyne detector, and transducer added noise $N_\textrm{t}$:
\begin{equation}
    \langle \hat{\zeta}_\textrm{I}(t) \hat{\zeta}_\textrm{I}(t')\rangle=\langle \hat{\zeta}_\textrm{Q}(t) \hat{\zeta}_\textrm{Q}(t')\rangle = \frac{1}{2}(1+N_\textrm{t})\delta(t-t').
    \label{eq:corramp}
\end{equation}

Eqs.~\eqref{eq:Iamp}, \eqref{eq:Qamp}, and \eqref{eq:corramp} define the amplifier model described in Fig.~\ref{fig:fig4}(b) of the main text in which the transduction process is characterised by an efficiency $\eta_\textrm{loss}$ and total two-quadrature noise $N_\textrm{det} = 1+N_\textrm{t}$ at the input of an ideal optical detector. But for qubit readout it is often more convenient to combine these metrics into a single figure of merit.  To this end, we rescale Eqs.~\eqref{eq:Iamp}, \eqref{eq:Qamp} to obtain
\begin{equation}
I(t) = \sqrt{G_\textrm{o}'}\left(\sqrt{\frac{\kappa_\textrm{c}\eta_\textrm{loss}}{1+ N_\textrm{t}}}\textrm{Re}(\alpha_{|k\rangle}(t)) + \tilde{\zeta}_\textrm{I}(t)\right)\label{eq:Iquad}
\end{equation}
\begin{equation}
Q(t) = \sqrt{G_\textrm{o}'}\left(\sqrt{\frac{\kappa_\textrm{c}\eta_\textrm{loss}}{1 + N_\textrm{t}}}\textrm{Im}(\alpha_{|k\rangle}(t)) + \tilde{\zeta}_\textrm{Q}(t)\right)\label{eq:Qquad},
\end{equation}
where $\sqrt{G_\textrm{o}'} = \sqrt{G_\textrm{o}(1+N_\textrm{t})}$ is a modified overall gain factor, and $\tilde{\zeta_\textrm{I}}$ and $\tilde{\zeta_\textrm{Q}}$ are noise operators containing only contributions from vacuum fluctuations and ideal heterodyne detection: $\langle \tilde{\zeta}_\textrm{I}(t)\tilde{\zeta}_\textrm{I}(t')\rangle = \langle \tilde{\zeta}_\textrm{Q}(t)\tilde{\zeta}_\textrm{Q}(t')\rangle =\frac{1}{2}\delta(t-t')$.  From Eqs.~\eqref{eq:Iquad} and \eqref{eq:Qquad} it is clear that $\eta_\textrm{q}$ can be defined as
\begin{equation}
    \eta_\textrm{q} = \eta_\textrm{loss}\eta_\textrm{noise} = \eta_\textrm{loss}\times\frac{1}{1+N_\textrm{t}}.
\end{equation}
See Supplementary Information for a derivation of the exact form of $N_\textrm{t}$.

\section{Prospects for quantum transduction}
The relevant figure of merit for quantum state transduction is the added noise referred to the output of the circuit QED system $N_\textrm{cQED} = N_\textrm{t}/\eta_\textrm{loss}$ \cite{zeuthen2020figures}.  When we optimise for maximum efficiency, with $(\Gamma_\textrm{e}, \Gamma_\textrm{o})/2\pi = (1.1, 5.0)$~kHz, and achieve $\eta_\textrm{q}\approx 8\times 10^{-4}$, we obtain $N_\textrm{cQED} = 740$~photons.  This is of course far from the value required for quantum-enabled transduction \cite{zeuthen2020figures} and is limited by insufficient bandwidth, LC circuit loss, transducer efficiency, and pump power-dependent noise generated by the LC circuit \cite{ground}.  However, even moderate improvements in the vacuum electromechanical coupling $g_\textrm{e}$ will greatly improve $\eta_\textrm{bw}$, $\eta_\textrm{t}$, and $\eta_\textrm{noise}$.  

An increase in $g_\textrm{e}$ would  provide many simultaneous improvements to the operation of the transducer by reducing the number of microwave pump photons required to achieve a given transduction bandwidth.  For example, $g_\textrm{e} = 2\pi \times 8$~Hz would enable $\Gamma_\textrm{e} = \Gamma_\textrm{o} = 2\pi\times 5$~kHz and reduce pump power-dependent LC circuit loss, yielding $\eta_\textrm{t}\approx0.6$.  If accompanied by an increase in the qubit decay times to $T_2 = T_1 = 100$~\si{\micro\second}, we expect $\eta_\textrm{bw} = 0.9$, and the transducer to be quite close to quantum-enabled with $N_\textrm{cQED} <10$.  Further reductions in added noise, which will be needed to approach the $N_\textrm{cQED}<1$ regime, are currently being explored, including the use of wafer bonding to achieve more reliable and even larger electromechanical coupling rates.  Pump-induced vortex loss in the superconductor is also currently being investigated as a possible source of both LC circuit loss and added noise, which may be improved through better magnetic shielding.

\section{Isolation and filtering}
The two modular systems are connected via a coaxial transmission line through a single-junction circulator in series with a triple-junction circulator.  This provides a total of 63~dB of isolation to shield the circuit QED system from the strong microwave pump (up to $-45$~dBm incident on the LC circuit), which is routed to the transducer through a directional coupler.  Additional isolation is achieved by sending a cancellation tone through the second port of the directional coupler (labelled ``cancel port'' in Extended Data Fig.~\ref{fig:backaction_ext}(a)) to interferometrically cancel the large reflected microwave pump \cite{mallet2011quantum}.  Filters are also placed between the circuit QED system and the transducer to eliminate the propagation of high-frequency thermal radiation along the coaxial cable connecting the two systems.  See Extended Data Fig.~\ref{fig:detailedDiagram}(c) for details of the layout and filtering. 

\section{Source of excess backaction}
As shown in Fig.~\ref{fig:backaction}(b) of the main text, there is a small amount of excess backaction from the operation of the transducer at very high $\Gamma_\textrm{e}$, equivalent to $\Delta n = (3\pm1)\times10^{-3}$ excess photons in the circuit QED system's microwave cavity.  Two possible mechanisms could cause such excess backaction: heating of the 50~$\Omega$ termination that thermalises the circuit QED system by the strong microwave pump, or direct backaction from pump photons making it through the isolation between the circuit QED system and the transducer.  We can differentiate these two possible sources of backaction by using our cancellation scheme as an interferometer to tune the power $P_\textrm{ref}$ of the microwave field propagating towards the circuit QED system (see Extended Data Fig.~\ref{fig:backaction_ext}(a),(b)).

The number $\bar{n}_\textrm{p}$ of pump photons in the circuit QED system's microwave cavity can be accurately measured by monitoring the qubit frequency shift $\Delta \omega_\textrm{q} = 2\chi \bar{n}_\textrm{p}$ due to the (coherent) AC Stark effect \cite{gambetta2006qubit}.  Extended Data Fig.~\ref{fig:backaction_ext}(c) shows the number of coherent pump photons in the circuit QED system's cavity as a function of $P_\textrm{ref}$.  During normal operation of the transducer the cancellation is optimised so that $P_\textrm{ref}<-95$~dBm, and we can linearly extrapolate $\bar{n}_\textrm{p}$ to low powers to find that $\bar{n}_\textrm{p}<1\times 10^{-3}$ during normal operation.  The measured excess backaction $\Delta n = (3\pm1)\times10^{-3}$ is larger, so it is likely that local heating of the 50~$\Omega$ termination that thermalises the circuit QED system is responsible for the majority of the excess backaction observed in Fig.~\ref{fig:backaction}(b) in the main text.  Improvements in $g_\textrm{e}$ that will enable a reduction in microwave pump power will serve to further reduce this excess backaction.  

\section{acknowledgements}
We acknowledge funding from AFOSR MURI Grant
No. FA9550-15-1-0015, from ARO CQTS Grant No. 67C1098620, and NSF under Grant No. PHYS 1734006.  We thank Jim Beall and Katarina Cicak for help with our fabrication process, as well as Ramin Lalezari for fabricating the optical cavity mirror coatings.  We thank John Teufel, Graeme Smith, Kyle Quinlan, Kazemi Adachi and Luca Talamo for feedback on the manuscript.    

\section{Supplementary Information}
Supplementary Information is available for this paper.
\section{author contributions}
R.D.D., C.A.R. and K.W.L. conceived of the experiment.  B.M.B., M.D.U., J.M.K., S.M. and R.D.D. planned and carried out the measurements.  M.D.U. constructed the optical cavity.  R.D.D. designed and fabricated the circuit QED system.  S.M. and P.S.B. developed the fabrication process for the chips hosting the electrical circuit and membrane, which were fabricated by S.M.  All authors contributed to writing the manuscript.    
\section{Data availability}
The experimental data and numerical calculations are available from the corresponding author upon reasonable request.


%\bibliography{references}
\FloatBarrier
\onecolumngrid
\setcounter{figure}{0}
\renewcommand{\figurename}{\textbf{Extended Data Fig.}}
\begin{figure*}
    \centering
    \includegraphics{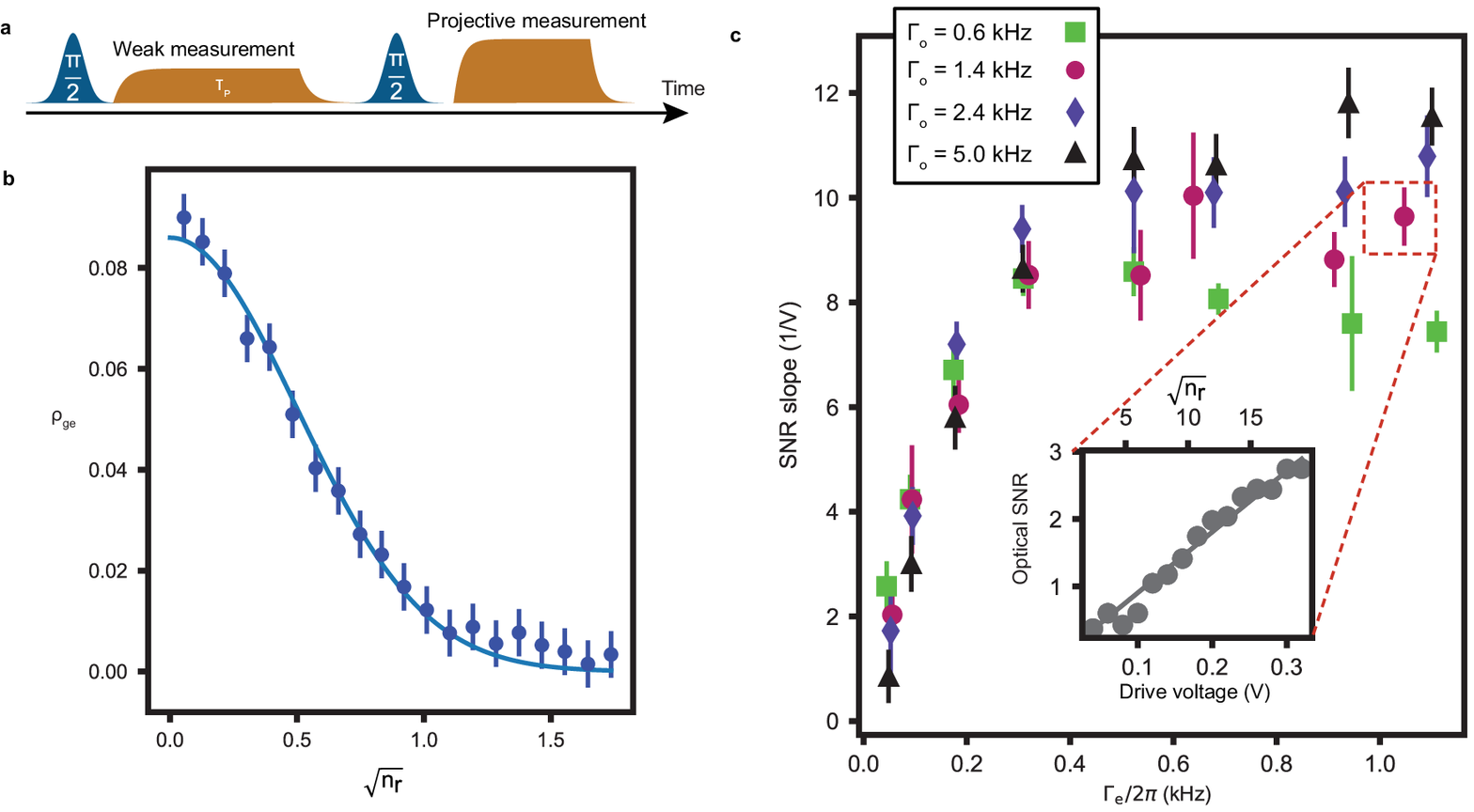}
    \caption{\textbf{Quantum efficiency measurement} (a)  Protocol for measurement-induced dephasing calibration.  A weak measurement pulse is injected into a Ramsey sequence.  The Ramsey sequence is then followed by a strong projective measurement of the qubit.  (b) The coherence of the qubit $\rho_\textrm{ge}$ decays as a Gaussian function of the weak measurement amplitude $\sqrt{\bar{n}_\textrm{r}}$.  The points are data, while the line is a Gaussian fit.  (c) Measurement of the SNR slope $a$ of the optically mediated readout, in units of inverse drive voltage, as a function of $\Gamma_\textrm{e}$ for four fixed optomechanical damping rates $\Gamma_\textrm{o}$.  Each data point is obtained from a fit to a measurement of the sort shown in the inset, where it is seen that the SNR scales linearly as a function of drive voltage (bottom axis) or equivalently readout pulse amplitude (top axis).} 
    \label{fig:quant_eff_ext}
\end{figure*}

\begin{figure*}
    \centering
    \includegraphics{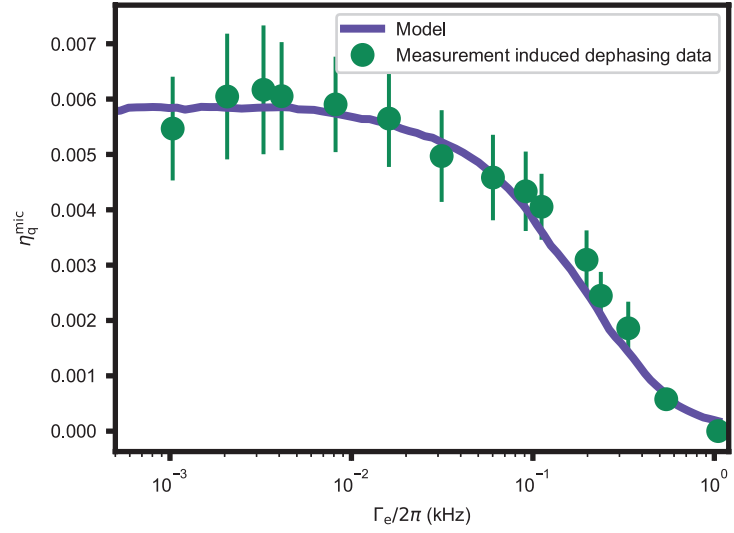}
    \caption{\textbf{Characterisation of the quantum efficiency of the microwave readout apparatus}.  The microwave readout efficiency $\eta_\textrm{q}^\textrm{mic}$ is measured as a function of the electromechanical damping rate $\Gamma_\textrm{e}$.  The points are data, while the line is a model including partial absorption of the readout pulse by the pump power-dependent reflection loss of the LC circuit, power-dependent added noise, fixed transmission losses and the independently measured added noise of the microwave heterodyne measurement chain.  The shape of the curve is dominated by power-dependent LC circuit reflection loss, with the quantum efficiency of the microwave readout apparatus approaching zero at high power because the LC circuit is nearly critically coupled. From this model, we estimate $4.7$~dB of loss ($\eta_\textrm{mic} = 0.34$)  between transducer and the circuit QED system.}
    \label{fig:mic_loss}
\end{figure*}

\begin{figure*}
    \centering
    \includegraphics{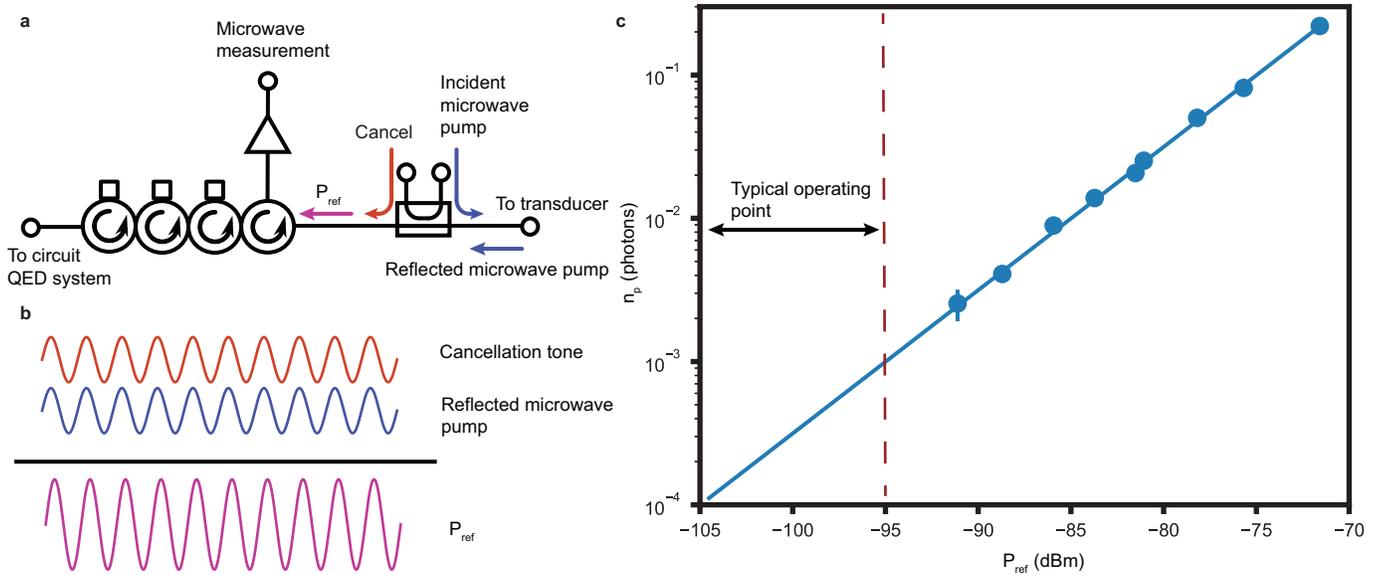}
    \caption{\textbf{Excess backaction from pump photon leakage} (a)   Circulators (total isolation of 63 dB) and interferometric cancellation are used to prevent the strong microwave pump from reaching the circuit QED system and dephasing the qubit.  Cancellation is achieved by sending a microwave cancellation tone with amplitude equal to that of the reflected microwave pump but opposite phase into the second arm of the directional coupler.  The phasor sum of this cancellation tone and the reflected microwave pump determines the pump power $P_\textrm{ref}$ propagating towards the circuit QED system.  (b) The directional coupler acts as an interferometer, enabling interference between the reflected microwave pump and the cancellation tone.  During normal operation of the transducer the cancellation tone is tuned to minimise $P_\textrm{ref}$, but to estimate the effect of pump photons on the qubit the cancellation tone can be tuned to only partially interfere and tune $P_\textrm{ref}$ over several decades.  The phasor sum shown here is for the case of constructive interference.  (c) The number of pump photons in the circuit QED system $\bar{n}_\textrm{p}$ is measured as $P_\textrm{ref}$ is varied.  During transducer operation the cancellation is typically tuned to achieve $P_\textrm{ref}<-95$~dBm or equivalently $\bar{n}_\textrm{p}<1\times10^{-3}$.  The cancellation does not stay completely fixed over the course of the experiment due to small thermal drifts slightly changing the amplitude of $P_\textrm{ref}$.}
    \label{fig:backaction_ext}
\end{figure*}
\clearpage
\begin{figure*}[t]
    \includegraphics{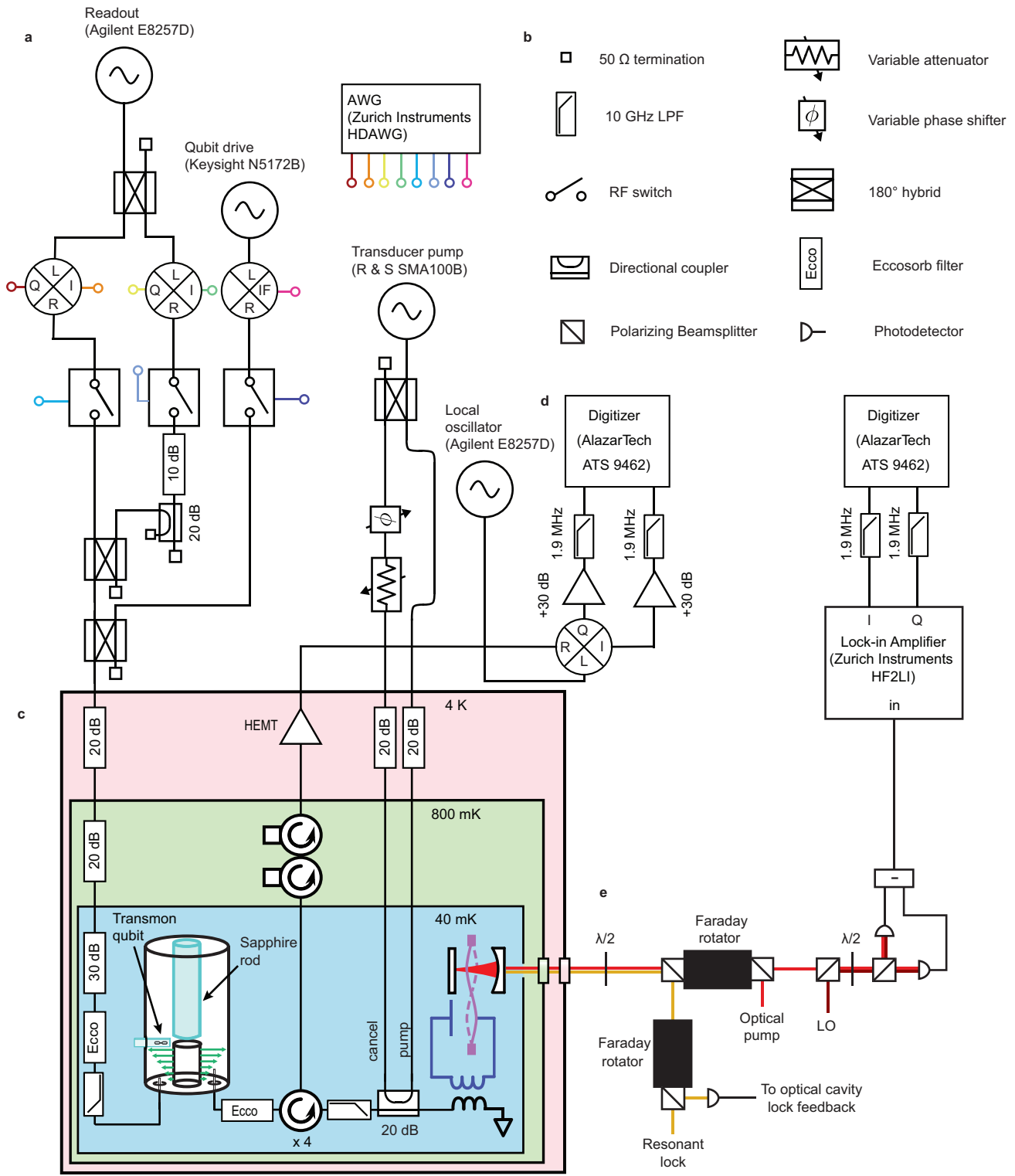}
    \caption{\textbf{Experimental Schematic}  (a) Microwave layout demonstrating the exact configuration of the qubit readout/control pulses and the pumps for the electro-optic transducer.  (b) Legend for various different microwave and optical components.  (c) Cryogenic portion of the experiment.  (d) Demodulation and detection scheme.  The two digitisers allow for simultaneous measurement of the microwave and optical signals emitted from the transducer.  (e) Simplified schematic of optical beam layout and balanced heterodyne detector.}
    \label{fig:detailedDiagram}
\end{figure*}

%\clearpage
%\begin{table}[h!]
%\caption{Circuit QED system parameters} 
%  \begin{center}
%    \begin{tabular}{ |p{3.3cm}|p{1.2cm} |p{3.5cm}| }
%    \hline
%    \textbf{Parameter} & \textbf{Symbol} & \textbf{Value} \\ \hline
%    Qubit frequency & $\omega_\textrm{q}$ & $\omega_\textrm{q}/2\pi = 5.632$~GHz \\ \hline
%    Qubit-cavity coupling & $g_\textrm{qc}$ & $g_\textrm{qc}/2\pi = 66.4$~MHz  \\ \hline
%    Qubit anharmonicity & $\nu$ & $\nu/2\pi = 228$~MHz \\ \hline
%    Dispersive shift & $\chi$ & $\chi/2\pi = 172$~kHz \\ \hline
%    Cavity frequency & $\omega_\textrm{c}$ & $\omega_\textrm{c}/2\pi = 7.938$~GHz \\ \hline
%    Cavity linewidth & $\kappa_\textrm{c}$ & $\kappa_\textrm{c}/2\pi = 380$~kHz \\ \hline
%    Weak port coupling & $\kappa_\textrm{c,w}$ & $\kappa_\textrm{c,w}/2\pi < 5$~kHz \\ \hline
%    Cavity internal loss & $\kappa_\textrm{c,int}$ & $\kappa_\textrm{c,int}/2\pi < 10$~kHz \\ \hline
%    Qubit lifetime & $T_1$ & $T_1 = 17$~\si{\micro\second} \\ \hline
%    Ramsey time & $T_2$ & $T_2 = 20$~\si{\micro\second} \\ \hline
%    \end{tabular}
%  \end{center}
%  \label{tab:circuitQED}
%\end{table}
\clearpage

\begin{table}[h!]
\caption{Circuit QED system and electro-optic transducer parameters.} 
  \begin{center}
    \begin{tabular}{ |p{7.5cm}|p{1.5cm} |p{3.9cm}| }
    \hline
    \textbf{Parameter} & \textbf{Symbol} & \textbf{Value} \\ \hline
    Qubit frequency & $\omega_\textrm{q}$ & $\omega_\textrm{q}/2\pi = 5.632$~GHz \\ \hline
    Qubit-cavity coupling & $g_\textrm{qc}$ & $g_\textrm{qc}/2\pi = 66.4$~MHz  \\ \hline
    Qubit anharmonicity & $\nu$ & $\nu/2\pi = 228$~MHz \\ \hline
    Dispersive shift & $\chi$ & $\chi/2\pi = 172$~kHz \\ \hline
    Cavity frequency & $\omega_\textrm{c}$ & $\omega_\textrm{c}/2\pi = 7.938$~GHz \\ \hline
    Cavity linewidth & $\kappa_\textrm{c}$ & $\kappa_\textrm{c}/2\pi = 380$~kHz \\ \hline
    Weak port coupling & $\kappa_\textrm{c,w}$ & $\kappa_\textrm{c,w}/2\pi < 5$~kHz \\ \hline
    Cavity internal loss & $\kappa_\textrm{c,int}$ & $\kappa_\textrm{c,int}/2\pi < 10$~kHz \\ \hline
    Qubit lifetime & $T_1$ & $T_1 = 17$~\si{\micro\second} \\ \hline
    Ramsey time & $T_2$ & $T_2 = 20$~\si{\micro\second} \\ \hline
    Optical cavity frequency &
    $\omega_\textrm{o}$ & $\omega_\textrm{o}/2\pi = 277 $~THz \\ \hline 

    Optical cavity external coupling & $\kappa_\textrm{o,ext}$ & $\kappa_\textrm{o,ext}/2\pi = 2.12$~MHz \\ \hline
    
    Optical cavity linewidth &
    $\kappa_\textrm{o}$ & $\kappa_\textrm{o}/2\pi = 2.68$~MHz \\ \hline
    
    LC circuit frequency &
    $\omega_\textrm{e}$ & $\omega_\textrm{e}/2\pi = 7.938 $~GHz \\ \hline
    
    LC circuit external coupling &
    $\kappa_\textrm{e,ext}$ & 
    $\kappa_\textrm{e,ext}/2\pi = 1.42$~MHz \\ \hline
    
    LC circuit linewidth (low pump power)&
    $\kappa_\textrm{e}$ & 
    $\kappa_\textrm{e}/2\pi \approx 1.6$~MHz \\ \hline
    
    LC circuit linewidth (high pump power) &
    $\kappa_\textrm{e}$ & 
    $\kappa_\textrm{e}/2\pi \approx 2.7$~MHz \\ \hline

    Mechanical mode frequency &
    $\omega_\textrm{m}$ & $\omega_\textrm{m}/2\pi = 1.45 $~MHz \\ \hline
    
    Intrinsic mechanical dissipation rate &
    $\gamma_\textrm{m}$ & $\gamma_\textrm{m}/2\pi = 0.11 $~Hz \\ \hline 
    
    Vacuum optomechanical coupling &
    $g_\textrm{o}$ & $g_\textrm{o}/2\pi = 60$~Hz \\ \hline
    
    Vacuum electromechanical coupling &
    $g_\textrm{e}$ & $g_\textrm{e}/2\pi = 1.6$~Hz \\ \hline
    
    Optical cavity mode matching & $\epsilon$ & $\epsilon = 0.80$ \\ \hline

    \end{tabular}
  \end{center}
  \label{tab:transducer}
\end{table}
\clearpage

\begin{table}[h!]
\caption{Contributions to the quantum efficiency. Numerical values are not reported for contributions that depend on $\Gamma_\textrm{e}$ or $\Gamma_\textrm{o}$.} 
  \begin{center}
    \begin{tabular}{ |p{6cm}|p{1.5cm} |p{5cm}| }
    \hline
    \textbf{Parameter} & \textbf{Symbol} & \textbf{Value} \\ \hline
    Finite bandwidth & $\eta_\textrm{bw}$ & see Eq.~\eqref{eq:eta_bw} \\ \hline

    Microwave transmission loss & $\eta_\textrm{mic}$ & $\eta_\textrm{mic} = 0.34$ \\ \hline
    
    Circuit QED system cavity loss & $\eta_\textrm{cav}$ & $\eta_\textrm{cav} >0.96$ \\ \hline
    
   Optical detection efficiency & $\eta_\textrm{opt}$ & $\eta_\textrm{opt} = 0.28$ \\ \hline
   
     Added noise & $\eta_\textrm{noise}$ & $\eta_\textrm{noise} = (1 + N_\textrm{t})^{-1} = N_\textrm{det}^{-1}$ \\ \hline 
  Transducer efficiency & $\eta_\textrm{t}$ & See Eq.~\eqref{eq:eta_conv} \\ \hline
   Transducer gain & $\eta_\textrm{G}$ & See Eq.~\eqref{eq:transd_gain} \\ \hline 
    \end{tabular}
      \label{tab:quantum_eff}
  \end{center}
\end{table}
%\bibliography{referencesBIB}

\begin{thebibliography}{0}%
\makeatletter
\providecommand \@ifxundefined [1]{%
 \@ifx{#1\undefined}
}%
\providecommand \@ifnum [1]{%
 \ifnum #1\expandafter \@firstoftwo
 \else \expandafter \@secondoftwo
 \fi
}%
\providecommand \@ifx [1]{%
 \ifx #1\expandafter \@firstoftwo
 \else \expandafter \@secondoftwo
 \fi
}%
\providecommand \natexlab [1]{#1}%
\providecommand \enquote  [1]{``#1''}%
\providecommand \bibnamefont  [1]{#1}%
\providecommand \bibfnamefont [1]{#1}%
\providecommand \citenamefont [1]{#1}%
\providecommand \href@noop [0]{\@secondoftwo}%
\providecommand \href [0]{\begingroup \@sanitize@url \@href}%
\providecommand \@href[1]{\@@startlink{#1}\@@href}%
\providecommand \@@href[1]{\endgroup#1\@@endlink}%
\providecommand \@sanitize@url [0]{\catcode `\\12\catcode `\$12\catcode
  `\&12\catcode `\#12\catcode `\^12\catcode `\_12\catcode `\%12\relax}%
\providecommand \@@startlink[1]{}%
\providecommand \@@endlink[0]{}%
\providecommand \url  [0]{\begingroup\@sanitize@url \@url }%
\providecommand \@url [1]{\endgroup\@href {#1}{\urlprefix }}%
\providecommand \urlprefix  [0]{URL }%
\providecommand \Eprint [0]{\href }%
\providecommand \doibase [0]{https://doi.org/}%
\providecommand \selectlanguage [0]{\@gobble}%
\providecommand \bibinfo  [0]{\@secondoftwo}%
\providecommand \bibfield  [0]{\@secondoftwo}%
\providecommand \translation [1]{[#1]}%
\providecommand \BibitemOpen [0]{}%
\providecommand \bibitemStop [0]{}%
\providecommand \bibitemNoStop [0]{.\EOS\space}%
\providecommand \EOS [0]{\spacefactor3000\relax}%
\providecommand \BibitemShut  [1]{\csname bibitem#1\endcsname}%
\let\auto@bib@innerbib\@empty
%</preamble>
\end{thebibliography}%


\begin{thebibliography}{41}%
\makeatletter
\providecommand \@ifxundefined [1]{%
 \@ifx{#1\undefined}
}%
\providecommand \@ifnum [1]{%
 \ifnum #1\expandafter \@firstoftwo
 \else \expandafter \@secondoftwo
 \fi
}%
\providecommand \@ifx [1]{%
 \ifx #1\expandafter \@firstoftwo
 \else \expandafter \@secondoftwo
 \fi
}%
\providecommand \natexlab [1]{#1}%
\providecommand \enquote  [1]{``#1''}%
\providecommand \bibnamefont  [1]{#1}%
\providecommand \bibfnamefont [1]{#1}%
\providecommand \citenamefont [1]{#1}%
\providecommand \href@noop [0]{\@secondoftwo}%
\providecommand \href [0]{\begingroup \@sanitize@url \@href}%
\providecommand \@href[1]{\@@startlink{#1}\@@href}%
\providecommand \@@href[1]{\endgroup#1\@@endlink}%
\providecommand \@sanitize@url [0]{\catcode `\\12\catcode `\$12\catcode
  `\&12\catcode `\#12\catcode `\^12\catcode `\_12\catcode `\%12\relax}%
\providecommand \@@startlink[1]{}%
\providecommand \@@endlink[0]{}%
\providecommand \url  [0]{\begingroup\@sanitize@url \@url }%
\providecommand \@url [1]{\endgroup\@href {#1}{\urlprefix }}%
\providecommand \urlprefix  [0]{URL }%
\providecommand \Eprint [0]{\href }%
\providecommand \doibase [0]{https://doi.org/}%
\providecommand \selectlanguage [0]{\@gobble}%
\providecommand \bibinfo  [0]{\@secondoftwo}%
\providecommand \bibfield  [0]{\@secondoftwo}%
\providecommand \translation [1]{[#1]}%
\providecommand \BibitemOpen [0]{}%
\providecommand \bibitemStop [0]{}%
\providecommand \bibitemNoStop [0]{.\EOS\space}%
\providecommand \EOS [0]{\spacefactor3000\relax}%
\providecommand \BibitemShut  [1]{\csname bibitem#1\endcsname}%
\let\auto@bib@innerbib\@empty
%</preamble>
\bibitem [{\citenamefont {Arute}\ \emph {et~al.}(2019)\citenamefont {Arute}
  \emph {et~al.}}]{arute2019quantum}%
  \BibitemOpen
  \bibfield  {author} {\bibinfo {author} {\bibfnamefont {F.}~\bibnamefont
  {Arute}} \emph {et~al.},\ }\bibfield  {title} {\bibinfo {title} {Quantum
  supremacy using a programmable superconducting processor},\ }\href@noop {}
  {\bibfield  {journal} {\bibinfo  {journal} {Nature}\ }\textbf {\bibinfo
  {volume} {574}},\ \bibinfo {pages} {505} (\bibinfo {year}
  {2019})}\BibitemShut {NoStop}%
\bibitem [{\citenamefont {Hisatomi}\ \emph {et~al.}(2016)\citenamefont
  {Hisatomi} \emph {et~al.}}]{hisatomi2016bidirectional}%
  \BibitemOpen
  \bibfield  {author} {\bibinfo {author} {\bibfnamefont {R.}~\bibnamefont
  {Hisatomi}} \emph {et~al.},\ }\bibfield  {title} {\bibinfo {title}
  {Bidirectional conversion between microwave and light via ferromagnetic
  magnons},\ }\href@noop {} {\bibfield  {journal} {\bibinfo  {journal} {Phys.
  Rev. B}\ }\textbf {\bibinfo {volume} {93}},\ \bibinfo {pages} {174427}
  (\bibinfo {year} {2016})}\BibitemShut {NoStop}%
\bibitem [{\citenamefont {Han}\ \emph {et~al.}(2018)\citenamefont {Han} \emph
  {et~al.}}]{han2018coherent}%
  \BibitemOpen
  \bibfield  {author} {\bibinfo {author} {\bibfnamefont {J.}~\bibnamefont
  {Han}} \emph {et~al.},\ }\bibfield  {title} {\bibinfo {title} {Coherent
  microwave-to-optical conversion via six-wave mixing in {Rydberg} atoms},\
  }\href@noop {} {\bibfield  {journal} {\bibinfo  {journal} {Phys. Rev. Lett.}\
  }\textbf {\bibinfo {volume} {120}},\ \bibinfo {pages} {093201} (\bibinfo
  {year} {2018})}\BibitemShut {NoStop}%
\bibitem [{\citenamefont {Higginbotham}\ \emph {et~al.}(2018)\citenamefont
  {Higginbotham} \emph {et~al.}}]{higginbotham2018harnessing}%
  \BibitemOpen
  \bibfield  {author} {\bibinfo {author} {\bibfnamefont {A.~P.}\ \bibnamefont
  {Higginbotham}} \emph {et~al.},\ }\bibfield  {title} {\bibinfo {title}
  {Harnessing electro-optic correlations in an efficient mechanical
  converter},\ }\href@noop {} {\bibfield  {journal} {\bibinfo  {journal} {Nat.
  Phys.}\ }\textbf {\bibinfo {volume} {14}},\ \bibinfo {pages} {1038} (\bibinfo
  {year} {2018})}\BibitemShut {NoStop}%
\bibitem [{\citenamefont {Bartholomew}\ \emph {et~al.}(2020)\citenamefont
  {Bartholomew} \emph {et~al.}}]{bartholomew2020chip}%
  \BibitemOpen
  \bibfield  {author} {\bibinfo {author} {\bibfnamefont {J.~G.}\ \bibnamefont
  {Bartholomew}} \emph {et~al.},\ }\bibfield  {title} {\bibinfo {title}
  {On-chip coherent microwave-to-optical transduction mediated by ytterbium in
  {YVO}$_4$},\ }\href@noop {} {\bibfield  {journal} {\bibinfo  {journal} {Nat.
  Commun.}\ }\textbf {\bibinfo {volume} {11}},\ \bibinfo {pages} {1} (\bibinfo
  {year} {2020})}\BibitemShut {NoStop}%
\bibitem [{\citenamefont {Stockill}\ \emph {et~al.}()\citenamefont {Stockill}
  \emph {et~al.}}]{stockill2021ultra-low-noise}%
  \BibitemOpen
  \bibfield  {author} {\bibinfo {author} {\bibfnamefont {R.}~\bibnamefont
  {Stockill}} \emph {et~al.},\ }\bibfield  {title} {\bibinfo {title}
  {Ultra-low-noise microwave to optics conversion in gallium phosphide},\
  }\href@noop {} {\bibinfo  {journal} {arXiv preprint arXiv:2107.04433}\
  }\BibitemShut {NoStop}%
\bibitem [{\citenamefont {Sahu}\ \emph {et~al.}()\citenamefont {Sahu} \emph
  {et~al.}}]{sahu2021quantum-enabled}%
  \BibitemOpen
\bibfield  {journal} {  }\bibfield  {author} {\bibinfo {author} {\bibfnamefont
  {R.}~\bibnamefont {Sahu}} \emph {et~al.},\ }\bibfield  {title} {\bibinfo
  {title} {Quantum-enabled interface between microwave and telecom light},\
  }\href@noop {} {\bibinfo  {journal} {arXiv preprint arXiv:2107.08303}\
  }\BibitemShut {NoStop}%
\bibitem [{\citenamefont {Barends}\ \emph {et~al.}(2011)\citenamefont {Barends}
  \emph {et~al.}}]{barends2011minimizing}%
  \BibitemOpen
\bibfield  {journal} {  }\bibfield  {author} {\bibinfo {author} {\bibfnamefont
  {R.}~\bibnamefont {Barends}} \emph {et~al.},\ }\bibfield  {title} {\bibinfo
  {title} {Minimizing quasiparticle generation from stray infrared light in
  superconducting quantum circuits},\ }\href@noop {} {\bibfield  {journal}
  {\bibinfo  {journal} {Appl. Phys. Lett.}\ }\textbf {\bibinfo {volume} {99}},\
  \bibinfo {pages} {113507} (\bibinfo {year} {2011})}\BibitemShut {NoStop}%
\bibitem [{\citenamefont {Mirhosseini}\ \emph {et~al.}(2020)\citenamefont
  {Mirhosseini}, \citenamefont {Sipahigil}, \citenamefont {Kalaee},\ and\
  \citenamefont {Painter}}]{mirhosseini2020superconducting}%
  \BibitemOpen
  \bibfield  {author} {\bibinfo {author} {\bibfnamefont {M.}~\bibnamefont
  {Mirhosseini}}, \bibinfo {author} {\bibfnamefont {A.}~\bibnamefont
  {Sipahigil}}, \bibinfo {author} {\bibfnamefont {M.}~\bibnamefont {Kalaee}},\
  and\ \bibinfo {author} {\bibfnamefont {O.}~\bibnamefont {Painter}},\
  }\bibfield  {title} {\bibinfo {title} {Superconducting qubit to optical
  photon transduction},\ }\href@noop {} {\bibfield  {journal} {\bibinfo
  {journal} {Nature}\ }\textbf {\bibinfo {volume} {588}},\ \bibinfo {pages}
  {599} (\bibinfo {year} {2020})}\BibitemShut {NoStop}%
\bibitem [{\citenamefont {Campbell}\ and\ \citenamefont
  {Benjamin}(2008)}]{campbell2008measurement}%
  \BibitemOpen
  \bibfield  {author} {\bibinfo {author} {\bibfnamefont {E.~T.}\ \bibnamefont
  {Campbell}}\ and\ \bibinfo {author} {\bibfnamefont {S.~C.}\ \bibnamefont
  {Benjamin}},\ }\bibfield  {title} {\bibinfo {title} {Measurement-based
  entanglement under conditions of extreme photon loss},\ }\href@noop {}
  {\bibfield  {journal} {\bibinfo  {journal} {Phys. Rev. Lett.}\ }\textbf
  {\bibinfo {volume} {101}},\ \bibinfo {pages} {130502} (\bibinfo {year}
  {2008})}\BibitemShut {NoStop}%
\bibitem [{\citenamefont {Kalb}\ \emph {et~al.}(2017)\citenamefont {Kalb} \emph
  {et~al.}}]{kalb2017entanglement}%
  \BibitemOpen
  \bibfield  {author} {\bibinfo {author} {\bibfnamefont {N.}~\bibnamefont
  {Kalb}} \emph {et~al.},\ }\bibfield  {title} {\bibinfo {title} {Entanglement
  distillation between solid-state quantum network nodes},\ }\href@noop {}
  {\bibfield  {journal} {\bibinfo  {journal} {Science}\ }\textbf {\bibinfo
  {volume} {356}},\ \bibinfo {pages} {928} (\bibinfo {year}
  {2017})}\BibitemShut {NoStop}%
\bibitem [{\citenamefont {Zhong}\ \emph {et~al.}(2020)\citenamefont {Zhong}
  \emph {et~al.}}]{zhong2020proposal}%
  \BibitemOpen
  \bibfield  {author} {\bibinfo {author} {\bibfnamefont {C.}~\bibnamefont
  {Zhong}} \emph {et~al.},\ }\bibfield  {title} {\bibinfo {title} {Proposal for
  heralded generation and detection of entangled microwave--optical-photon
  pairs},\ }\href@noop {} {\bibfield  {journal} {\bibinfo  {journal} {Phys.
  Rev. Lett.}\ }\textbf {\bibinfo {volume} {124}},\ \bibinfo {pages} {010511}
  (\bibinfo {year} {2020})}\BibitemShut {NoStop}%
\bibitem [{\citenamefont {Kurpiers}\ \emph {et~al.}(2019)\citenamefont
  {Kurpiers} \emph {et~al.}}]{kurpiers2019quantum}%
  \BibitemOpen
  \bibfield  {author} {\bibinfo {author} {\bibfnamefont {P.}~\bibnamefont
  {Kurpiers}} \emph {et~al.},\ }\bibfield  {title} {\bibinfo {title} {Quantum
  communication with time-bin encoded microwave photons},\ }\href@noop {}
  {\bibfield  {journal} {\bibinfo  {journal} {Phys. Rev. Appl.}\ }\textbf
  {\bibinfo {volume} {12}},\ \bibinfo {pages} {044067} (\bibinfo {year}
  {2019})}\BibitemShut {NoStop}%
\bibitem [{\citenamefont {Hatridge}\ \emph {et~al.}(2013)\citenamefont
  {Hatridge} \emph {et~al.}}]{hatridge2013quantum}%
  \BibitemOpen
  \bibfield  {author} {\bibinfo {author} {\bibfnamefont {M.}~\bibnamefont
  {Hatridge}} \emph {et~al.},\ }\bibfield  {title} {\bibinfo {title} {Quantum
  back-action of an individual variable-strength measurement},\ }\href@noop {}
  {\bibfield  {journal} {\bibinfo  {journal} {Science}\ }\textbf {\bibinfo
  {volume} {339}},\ \bibinfo {pages} {178} (\bibinfo {year}
  {2013})}\BibitemShut {NoStop}%
\bibitem [{\citenamefont {Inagaki}\ \emph {et~al.}(2013)\citenamefont {Inagaki}
  \emph {et~al.}}]{inagaki2013entanglement}%
  \BibitemOpen
  \bibfield  {author} {\bibinfo {author} {\bibfnamefont {T.}~\bibnamefont
  {Inagaki}} \emph {et~al.},\ }\bibfield  {title} {\bibinfo {title}
  {Entanglement distribution over 300 km of fiber},\ }\href@noop {} {\bibfield
  {journal} {\bibinfo  {journal} {Opt. Express}\ }\textbf {\bibinfo {volume}
  {21}},\ \bibinfo {pages} {23241} (\bibinfo {year} {2013})}\BibitemShut
  {NoStop}%
\bibitem [{\citenamefont {Ursin}\ \emph {et~al.}(2007)\citenamefont {Ursin}
  \emph {et~al.}}]{ursin2007entanglement}%
  \BibitemOpen
  \bibfield  {author} {\bibinfo {author} {\bibfnamefont {R.}~\bibnamefont
  {Ursin}} \emph {et~al.},\ }\bibfield  {title} {\bibinfo {title}
  {Entanglement-based quantum communication over 144 km},\ }\href@noop {}
  {\bibfield  {journal} {\bibinfo  {journal} {Nat. Phys.}\ }\textbf {\bibinfo
  {volume} {3}},\ \bibinfo {pages} {481} (\bibinfo {year} {2007})}\BibitemShut
  {NoStop}%
\bibitem [{\citenamefont {Yin}\ \emph {et~al.}(2017)\citenamefont {Yin} \emph
  {et~al.}}]{yin2017satellite}%
  \BibitemOpen
  \bibfield  {author} {\bibinfo {author} {\bibfnamefont {J.}~\bibnamefont
  {Yin}} \emph {et~al.},\ }\bibfield  {title} {\bibinfo {title}
  {Satellite-based entanglement distribution over 1200 kilometers},\
  }\href@noop {} {\bibfield  {journal} {\bibinfo  {journal} {Science}\ }\textbf
  {\bibinfo {volume} {356}},\ \bibinfo {pages} {1140} (\bibinfo {year}
  {2017})}\BibitemShut {NoStop}%
\bibitem [{\citenamefont {Lecocq}\ \emph {et~al.}(2021)\citenamefont {Lecocq}
  \emph {et~al.}}]{lecocq2021control}%
  \BibitemOpen
  \bibfield  {author} {\bibinfo {author} {\bibfnamefont {F.}~\bibnamefont
  {Lecocq}} \emph {et~al.},\ }\bibfield  {title} {\bibinfo {title} {Control and
  readout of a superconducting qubit using a photonic link},\ }\href@noop {}
  {\bibfield  {journal} {\bibinfo  {journal} {Nature}\ }\textbf {\bibinfo
  {volume} {591}},\ \bibinfo {pages} {575} (\bibinfo {year}
  {2021})}\BibitemShut {NoStop}%
\bibitem [{\citenamefont {Youssefi}\ \emph {et~al.}(2021)\citenamefont
  {Youssefi} \emph {et~al.}}]{youssefi2021cryogenic}%
  \BibitemOpen
  \bibfield  {author} {\bibinfo {author} {\bibfnamefont {A.}~\bibnamefont
  {Youssefi}} \emph {et~al.},\ }\bibfield  {title} {\bibinfo {title} {A
  cryogenic electro-optic interconnect for superconducting devices},\
  }\href@noop {} {\bibfield  {journal} {\bibinfo  {journal} {Nat. Electron.}\
  }\textbf {\bibinfo {volume} {4}},\ \bibinfo {pages} {326} (\bibinfo {year}
  {2021})}\BibitemShut {NoStop}%
\bibitem [{\citenamefont {Zeuthen}\ \emph {et~al.}(2020)\citenamefont
  {Zeuthen}, \citenamefont {Schliesser}, \citenamefont {S{\o}rensen},\ and\
  \citenamefont {Taylor}}]{zeuthen2020figures}%
  \BibitemOpen
  \bibfield  {author} {\bibinfo {author} {\bibfnamefont {E.}~\bibnamefont
  {Zeuthen}}, \bibinfo {author} {\bibfnamefont {A.}~\bibnamefont {Schliesser}},
  \bibinfo {author} {\bibfnamefont {A.~S.}\ \bibnamefont {S{\o}rensen}},\ and\
  \bibinfo {author} {\bibfnamefont {J.~M.}\ \bibnamefont {Taylor}},\ }\bibfield
   {title} {\bibinfo {title} {Figures of merit for quantum transducers},\
  }\href@noop {} {\bibfield  {journal} {\bibinfo  {journal} {Quantum Sci.
  Technol.}\ }\textbf {\bibinfo {volume} {5}},\ \bibinfo {pages} {034009}
  (\bibinfo {year} {2020})}\BibitemShut {NoStop}%
\bibitem [{\citenamefont {Reagor}\ \emph {et~al.}(2016)\citenamefont {Reagor}
  \emph {et~al.}}]{reagor2016quantum}%
  \BibitemOpen
  \bibfield  {author} {\bibinfo {author} {\bibfnamefont {M.}~\bibnamefont
  {Reagor}} \emph {et~al.},\ }\bibfield  {title} {\bibinfo {title} {Quantum
  memory with millisecond coherence in circuit {QED}},\ }\href@noop {}
  {\bibfield  {journal} {\bibinfo  {journal} {Phys. Rev. B}\ }\textbf {\bibinfo
  {volume} {94}},\ \bibinfo {pages} {014506} (\bibinfo {year}
  {2016})}\BibitemShut {NoStop}%
\bibitem [{\citenamefont {Campagne-Ibarcq}\ \emph {et~al.}(2020)\citenamefont
  {Campagne-Ibarcq} \emph {et~al.}}]{campagne2020quantum}%
  \BibitemOpen
  \bibfield  {author} {\bibinfo {author} {\bibfnamefont {P.}~\bibnamefont
  {Campagne-Ibarcq}} \emph {et~al.},\ }\bibfield  {title} {\bibinfo {title}
  {Quantum error correction of a qubit encoded in grid states of an
  oscillator},\ }\href@noop {} {\bibfield  {journal} {\bibinfo  {journal}
  {Nature}\ }\textbf {\bibinfo {volume} {584}},\ \bibinfo {pages} {368}
  (\bibinfo {year} {2020})}\BibitemShut {NoStop}%
\bibitem [{\citenamefont {Chakram}\ \emph {et~al.}(2021)\citenamefont {Chakram}
  \emph {et~al.}}]{chakram2021seamless}%
  \BibitemOpen
  \bibfield  {author} {\bibinfo {author} {\bibfnamefont {S.}~\bibnamefont
  {Chakram}} \emph {et~al.},\ }\bibfield  {title} {\bibinfo {title} {Seamless
  high-{Q} microwave cavities for multimode circuit quantum electrodynamics},\
  }\href@noop {} {\bibfield  {journal} {\bibinfo  {journal} {Physical Review
  Letters}\ }\textbf {\bibinfo {volume} {127}},\ \bibinfo {pages} {107701}
  (\bibinfo {year} {2021})}\BibitemShut {NoStop}%
\bibitem [{\citenamefont {Bultink}\ \emph {et~al.}(2018)\citenamefont {Bultink}
  \emph {et~al.}}]{bultink2018general}%
  \BibitemOpen
  \bibfield  {author} {\bibinfo {author} {\bibfnamefont {C.~C.}\ \bibnamefont
  {Bultink}} \emph {et~al.},\ }\bibfield  {title} {\bibinfo {title} {General
  method for extracting the quantum efficiency of dispersive qubit readout in
  circuit {QED}},\ }\href@noop {} {\bibfield  {journal} {\bibinfo  {journal}
  {Appl. Phys. Lett.}\ }\textbf {\bibinfo {volume} {112}},\ \bibinfo {pages}
  {092601} (\bibinfo {year} {2018})}\BibitemShut {NoStop}%
\bibitem [{\citenamefont {Clerk}\ \emph {et~al.}(2010)\citenamefont {Clerk}
  \emph {et~al.}}]{clerk2010introduction}%
  \BibitemOpen
  \bibfield  {author} {\bibinfo {author} {\bibfnamefont {A.~A.}\ \bibnamefont
  {Clerk}} \emph {et~al.},\ }\bibfield  {title} {\bibinfo {title} {Introduction
  to quantum noise, measurement, and amplification},\ }\href@noop {} {\bibfield
   {journal} {\bibinfo  {journal} {Rev. Mod. Phys.}\ }\textbf {\bibinfo
  {volume} {82}},\ \bibinfo {pages} {1155} (\bibinfo {year}
  {2010})}\BibitemShut {NoStop}%
\bibitem [{\citenamefont {Blais}\ \emph {et~al.}(2004)\citenamefont {Blais}
  \emph {et~al.}}]{blais2004cavity}%
  \BibitemOpen
  \bibfield  {author} {\bibinfo {author} {\bibfnamefont {A.}~\bibnamefont
  {Blais}} \emph {et~al.},\ }\bibfield  {title} {\bibinfo {title} {Cavity
  quantum electrodynamics for superconducting electrical circuits: An
  architecture for quantum computation},\ }\href@noop {} {\bibfield  {journal}
  {\bibinfo  {journal} {Phys. Rev. A}\ }\textbf {\bibinfo {volume} {69}},\
  \bibinfo {pages} {062320} (\bibinfo {year} {2004})}\BibitemShut {NoStop}%
\bibitem [{\citenamefont {Gambetta}\ \emph {et~al.}(2007)\citenamefont
  {Gambetta} \emph {et~al.}}]{gambetta2007protocols}%
  \BibitemOpen
  \bibfield  {author} {\bibinfo {author} {\bibfnamefont {J.~M.}\ \bibnamefont
  {Gambetta}} \emph {et~al.},\ }\bibfield  {title} {\bibinfo {title} {Protocols
  for optimal readout of qubits using a continuous quantum nondemolition
  measurement},\ }\href@noop {} {\bibfield  {journal} {\bibinfo  {journal}
  {Phys. Rev. A}\ }\textbf {\bibinfo {volume} {76}},\ \bibinfo {pages} {012325}
  (\bibinfo {year} {2007})}\BibitemShut {NoStop}%
\bibitem [{\citenamefont {Caves}(1982)}]{caves1982quantum}%
  \BibitemOpen
  \bibfield  {author} {\bibinfo {author} {\bibfnamefont {C.~M.}\ \bibnamefont
  {Caves}},\ }\bibfield  {title} {\bibinfo {title} {Quantum limits on noise in
  linear amplifiers},\ }\href@noop {} {\bibfield  {journal} {\bibinfo
  {journal} {Phys. Rev. D}\ }\textbf {\bibinfo {volume} {26}},\ \bibinfo
  {pages} {1817} (\bibinfo {year} {1982})}\BibitemShut {NoStop}%
\bibitem [{\citenamefont {Clerk}\ and\ \citenamefont
  {Utami}(2007)}]{clerk2007using}%
  \BibitemOpen
  \bibfield  {author} {\bibinfo {author} {\bibfnamefont {A.~A.}\ \bibnamefont
  {Clerk}}\ and\ \bibinfo {author} {\bibfnamefont {D.~W.}\ \bibnamefont
  {Utami}},\ }\bibfield  {title} {\bibinfo {title} {Using a qubit to measure
  photon-number statistics of a driven thermal oscillator},\ }\href@noop {}
  {\bibfield  {journal} {\bibinfo  {journal} {Phys. Rev. A}\ }\textbf {\bibinfo
  {volume} {75}},\ \bibinfo {pages} {042302} (\bibinfo {year}
  {2007})}\BibitemShut {NoStop}%
\bibitem [{\citenamefont {Mallet}\ \emph {et~al.}(2011)\citenamefont {Mallet}
  \emph {et~al.}}]{mallet2011quantum}%
  \BibitemOpen
  \bibfield  {author} {\bibinfo {author} {\bibfnamefont {F.}~\bibnamefont
  {Mallet}} \emph {et~al.},\ }\bibfield  {title} {\bibinfo {title} {Quantum
  state tomography of an itinerant squeezed microwave field},\ }\href@noop {}
  {\bibfield  {journal} {\bibinfo  {journal} {Phys. Rev. Lett.}\ }\textbf
  {\bibinfo {volume} {106}},\ \bibinfo {pages} {220502} (\bibinfo {year}
  {2011})}\BibitemShut {NoStop}%
\bibitem [{\citenamefont {Krantz}\ \emph {et~al.}(2019)\citenamefont {Krantz}
  \emph {et~al.}}]{krantz2019quantum}%
  \BibitemOpen
  \bibfield  {author} {\bibinfo {author} {\bibfnamefont {P.}~\bibnamefont
  {Krantz}} \emph {et~al.},\ }\bibfield  {title} {\bibinfo {title} {A quantum
  engineer's guide to superconducting qubits},\ }\href@noop {} {\bibfield
  {journal} {\bibinfo  {journal} {Appl. Phys. Rev.}\ }\textbf {\bibinfo
  {volume} {6}},\ \bibinfo {pages} {021318} (\bibinfo {year}
  {2019})}\BibitemShut {NoStop}%
\bibitem [{\citenamefont {Rosenthal}\ \emph {et~al.}(2021)\citenamefont
  {Rosenthal} \emph {et~al.}}]{rosenthal2021efficient}%
  \BibitemOpen
  \bibfield  {author} {\bibinfo {author} {\bibfnamefont {E.~I.}\ \bibnamefont
  {Rosenthal}} \emph {et~al.},\ }\bibfield  {title} {\bibinfo {title}
  {Efficient and low-backaction quantum measurement using a chip-scale
  detector},\ }\href@noop {} {\bibfield  {journal} {\bibinfo  {journal} {Phys.
  Rev. Lett.}\ }\textbf {\bibinfo {volume} {126}},\ \bibinfo {pages} {090503}
  (\bibinfo {year} {2021})}\BibitemShut {NoStop}%
\bibitem [{\citenamefont {Touzard}\ \emph {et~al.}(2019)\citenamefont {Touzard}
  \emph {et~al.}}]{touzard2019gated}%
  \BibitemOpen
  \bibfield  {author} {\bibinfo {author} {\bibfnamefont {S.}~\bibnamefont
  {Touzard}} \emph {et~al.},\ }\bibfield  {title} {\bibinfo {title} {Gated
  conditional displacement readout of superconducting qubits},\ }\href@noop {}
  {\bibfield  {journal} {\bibinfo  {journal} {Phys. Rev. Lett.}\ }\textbf
  {\bibinfo {volume} {122}},\ \bibinfo {pages} {080502} (\bibinfo {year}
  {2019})}\BibitemShut {NoStop}%
\bibitem [{\citenamefont {Gambetta}\ \emph {et~al.}(2006)\citenamefont
  {Gambetta} \emph {et~al.}}]{gambetta2006qubit}%
  \BibitemOpen
  \bibfield  {author} {\bibinfo {author} {\bibfnamefont {J.~M.}\ \bibnamefont
  {Gambetta}} \emph {et~al.},\ }\bibfield  {title} {\bibinfo {title}
  {Qubit-photon interactions in a cavity: Measurement-induced dephasing and
  number splitting},\ }\href@noop {} {\bibfield  {journal} {\bibinfo  {journal}
  {Phys. Rev. A}\ }\textbf {\bibinfo {volume} {74}},\ \bibinfo {pages} {042318}
  (\bibinfo {year} {2006})}\BibitemShut {NoStop}%
\bibitem [{\citenamefont {Brubaker}\ \emph {et~al.}()\citenamefont {Brubaker}
  \emph {et~al.}}]{ground}%
  \BibitemOpen
  \bibfield  {author} {\bibinfo {author} {\bibfnamefont {B.~M.}\ \bibnamefont
  {Brubaker}} \emph {et~al.},\ }\bibinfo {note} {Optomechanical ground-state cooling in a continuous and efficient electro-optict transducer, in preparation}\BibitemShut
  {NoStop}%
\bibitem [{\citenamefont {Place}\ \emph {et~al.}(2021)\citenamefont {Place}
  \emph {et~al.}}]{place2021new}%
  \BibitemOpen
  \bibfield  {author} {\bibinfo {author} {\bibfnamefont {A.~P.~M.}\
  \bibnamefont {Place}} \emph {et~al.},\ }\bibfield  {title} {\bibinfo {title}
  {New material platform for superconducting transmon qubits with coherence
  times exceeding 0.3 milliseconds},\ }\href@noop {} {\bibfield  {journal}
  {\bibinfo  {journal} {Nat. Commun.}\ }\textbf {\bibinfo {volume} {12}},\
  \bibinfo {pages} {1} (\bibinfo {year} {2021})}\BibitemShut {NoStop}%
\end{thebibliography}

\begin{thebibliography}{41}%
\makeatletter
\providecommand \@ifxundefined [1]{%
 \@ifx{#1\undefined}
}%
\providecommand \@ifnum [1]{%
 \ifnum #1\expandafter \@firstoftwo
 \else \expandafter \@secondoftwo
 \fi
}%
\providecommand \@ifx [1]{%
 \ifx #1\expandafter \@firstoftwo
 \else \expandafter \@secondoftwo
 \fi
}%
\providecommand \natexlab [1]{#1}%
\providecommand \enquote  [1]{``#1''}%
\providecommand \bibnamefont  [1]{#1}%
\providecommand \bibfnamefont [1]{#1}%
\providecommand \citenamefont [1]{#1}%
\providecommand \href@noop [0]{\@secondoftwo}%
\providecommand \href [0]{\begingroup \@sanitize@url \@href}%
\providecommand \@href[1]{\@@startlink{#1}\@@href}%
\providecommand \@@href[1]{\endgroup#1\@@endlink}%
\providecommand \@sanitize@url [0]{\catcode `\\12\catcode `\$12\catcode
  `\&12\catcode `\#12\catcode `\^12\catcode `\_12\catcode `\%12\relax}%
\providecommand \@@startlink[1]{}%
\providecommand \@@endlink[0]{}%
\providecommand \url  [0]{\begingroup\@sanitize@url \@url }%
\providecommand \@url [1]{\endgroup\@href {#1}{\urlprefix }}%
\providecommand \urlprefix  [0]{URL }%
\providecommand \Eprint [0]{\href }%
\providecommand \doibase [0]{https://doi.org/}%
\providecommand \selectlanguage [0]{\@gobble}%
\providecommand \bibinfo  [0]{\@secondoftwo}%
\providecommand \bibfield  [0]{\@secondoftwo}%
\providecommand \translation [1]{[#1]}%
\providecommand \BibitemOpen [0]{}%
\providecommand \bibitemStop [0]{}%
\providecommand \bibitemNoStop [0]{.\EOS\space}%
\providecommand \EOS [0]{\spacefactor3000\relax}%
\providecommand \BibitemShut  [1]{\csname bibitem#1\endcsname}%
\let\auto@bib@innerbib\@empty
\addtocounter{\@listctr}{36}
%</preamble>

\bibitem [{\citenamefont {Dolan}(1977)}]{dolan1977offset}%
  \BibitemOpen
  \bibfield  {author} {\bibinfo {author} {\bibfnamefont {G.~J.}\ \bibnamefont
  {Dolan}},\ }\bibfield  {title} {\bibinfo {title} {Offset masks for lift-off
  photoprocessing},\ }\href@noop {} {\bibfield  {journal} {\bibinfo  {journal}
  {Appl. Phys. Lett.}\ }\textbf {\bibinfo {volume} {31}},\ \bibinfo {pages}
  {337} (\bibinfo {year} {1977})}\BibitemShut {NoStop}%
\bibitem [{\citenamefont {Koch}\ \emph {et~al.}(2007)\citenamefont {Koch} \emph
  {et~al.}}]{koch2007charge}%
  \BibitemOpen
  \bibfield  {author} {\bibinfo {author} {\bibfnamefont {J.}~\bibnamefont
  {Koch}} \emph {et~al.},\ }\bibfield  {title} {\bibinfo {title}
  {Charge-insensitive qubit design derived from the {Cooper} pair box},\
  }\href@noop {} {\bibfield  {journal} {\bibinfo  {journal} {Phys. Rev. A}\
  }\textbf {\bibinfo {volume} {76}},\ \bibinfo {pages} {042319} (\bibinfo
  {year} {2007})}\BibitemShut {NoStop}%
\bibitem [{\citenamefont {Houck}\ \emph {et~al.}(2008)\citenamefont {Houck}
  \emph {et~al.}}]{houck2008controlling}%
  \BibitemOpen
  \bibfield  {author} {\bibinfo {author} {\bibfnamefont {A.~A.}\ \bibnamefont
  {Houck}} \emph {et~al.},\ }\bibfield  {title} {\bibinfo {title} {Controlling
  the spontaneous emission of a superconducting transmon qubit},\ }\href@noop
  {} {\bibfield  {journal} {\bibinfo  {journal} {Phys. Rev. Lett.}\
  }\textbf {\bibinfo {volume} {101}},\ \bibinfo {pages} {080502} (\bibinfo
  {year} {2008})}\BibitemShut {NoStop}%
\bibitem [{\citenamefont {Andrews}\ \emph {et~al.}(2014)\citenamefont {Andrews}
  \emph {et~al.}}]{andrews2014bidirectional}%
  \BibitemOpen
  \bibfield  {author} {\bibinfo {author} {\bibfnamefont {R.~W.}\ \bibnamefont
  {Andrews}} \emph {et~al.},\ }\bibfield  {title} {\bibinfo {title}
  {Bidirectional and efficient conversion between microwave and optical
  light},\ }\href@noop {} {\bibfield  {journal} {\bibinfo  {journal} {Nat.
  Phys.}\ }\textbf {\bibinfo {volume} {10}},\ \bibinfo {pages} {321} (\bibinfo
  {year} {2014})}\BibitemShut {NoStop}%
\bibitem [{\citenamefont {Aspelmeyer}\ \emph {et~al.}(2014)\citenamefont
  {Aspelmeyer}, \citenamefont {Kippenberg},\ and\ \citenamefont
  {Marquardt}}]{aspelmeyer2014cavity}%
  \BibitemOpen
  \bibfield  {author} {\bibinfo {author} {\bibfnamefont {M.}~\bibnamefont
  {Aspelmeyer}}, \bibinfo {author} {\bibfnamefont {T.~J.}\ \bibnamefont
  {Kippenberg}},\ and\ \bibinfo {author} {\bibfnamefont {F.}~\bibnamefont
  {Marquardt}},\ }\bibfield  {title} {\bibinfo {title} {Cavity optomechanics},\
  }\href@noop {} {\bibfield  {journal} {\bibinfo  {journal} {Rev. Mod. Phys.}\
  }\textbf {\bibinfo {volume} {86}},\ \bibinfo {pages} {1391} (\bibinfo {year}
  {2014})}\BibitemShut {NoStop}%
\end{thebibliography}
%TC:endignore
\end{document}

%\preprint{APS/123-QED}

\title{Supplementary material: Non-destructive optical readout of a superconducting qubit}% Force line breaks with \\
\maketitle   
\section{Optical readout fidelity}
The maximum optical readout fidelity of $F = F_\textrm{o}\,\text{erf}(\sqrt{2\eta_\textrm{q} n_\textrm{r}})  \approx 0.4$ is consistent with the measured quantum efficiency and a residual excited state population of $10-15\%$ in the qubit.  This residual occupancy is likely due to the relatively low attenuation ($51$~dB) on the microwave pump and cancellation lines between ambient temperature and the dilution refrigerator base plate  \cite{yan2018distinguishing}, which is required to deliver the high-power microwave pump to the electro-optic transducer.  However, further work is needed to determine whether other aspects of the optical-access cryostat contribute to this thermal population in the qubit.

\section{Experimental layout}\label{sub:layout}

A schematic of the experimental setup is shown in Extended Data Fig.~4. Here we briefly summarise the details of the optical setup, discussed further in Ref.~\cite{ground}. 

A beam sourced by a low-noise external cavity diode laser (Toptica Photonics CTL) operated at a wavelength $\lambda = 1084$~nm is passed through an optical filter cavity (not shown) to further reduce laser phase noise, and then split three ways to obtain pump (red), lock (yellow), and local oscillator (LO, maroon) beams. The pump and lock beams are frequency shifted relative to the LO beam by acousto-optic modulators, such that the lock beam is resonant with the optical cavity mode, the pump beam is red-detuned by $\omega_\textrm{m}$, and the LO beam is detuned from the pump beam by $12.8$~MHz to enable heterodyne measurement. The lock and pump beams are orthogonally polarised and thus can be combined and routed to separate detectors using a polarising beamsplitter.

The Pound-Drever-Hall technique is employed to simultaneously lock the laser to both the optical filter cavity and the transducer's optical cavity. The error signal is used to feed back to the laser wavelength, the filter cavity length, and the rf drive to an acousto-optic modulator. The absence of tuning elements on the transducer's optical cavity leads to improved stability relative to cavities in previous devices  \cite{higginbotham2018harnessing}.

As shown in Extended Data Fig.~4, the qubit control and readout pulses are generated via individual sources and controlled and gated by mixers, switches and an arbitrary waveform generator \cite{krantz2019quantum}.

\section{Modelling transducer added noise}  
To obtain an expression for the transducer added noise $N_\textrm{t}$, we begin with the spectral density of the noise at the input of the optical heterodyne detector, which we measure directly at each $(\Gamma_\textrm{e},\Gamma_\textrm{o})$ data point in Fig.~4(c) of the main text.  When normalised to the measured shot noise of the LO beam, the single-quadrature power spectral density in units of $\textrm{photons}/\textrm{s}/\textrm{Hz}$ is given by
\begin{equation}
    S_\textrm{out}(\omega) = \frac{1}{2}(1+S_\textrm{b}) + \frac{\Gamma_\textrm{T}^2}{4}\frac{S_\textrm{t}(0)}{\frac{\Gamma_\textrm{T}^2}{4} + \omega^2},
\end{equation}
where $S_\textrm{t}(0)$ is the amplitude of the Lorentzian frequency response on resonance.  The first term is the sum of equal contributions from vacuum fluctuations and the added noise of an ideal heterodyne detector, $S_\textrm{b}$ encodes a small contribution from phase noise on the optical pump, and the Lorentzian term is due to fluctuations in the motion of the membrane imprinted on the reflected optical pump \cite{ground, teufel2011sideband}. The noise operator autocorrelation functions are obtained from the inverse Fourier transform of this spectrum:
\begin{equation}
    \langle \hat{\zeta}_\textrm{I}(t)\hat{\zeta}_\textrm{I}(t')\rangle  =\frac{1}{2}\left(1+S_\textrm{b}\right)\delta(t-t') + \frac{\Gamma_\textrm{T}}{4}S_\textrm{t}(0)e^{-\Gamma_\textrm{T}|t-t'|/2},
\end{equation}
and $\langle \hat{\zeta}_\textrm{Q}(t)\hat{\zeta}_\textrm{Q}(t')\rangle = \langle \hat{\zeta}_\textrm{I}(t)\hat{\zeta}_\textrm{I}(t')\rangle$.

To obtain a single value encoding the state of the qubit, we perform a weighted integral of the two quadrature voltage records of the form \cite{bultink2018general}
\begin{equation}
    V_{|k\rangle}(T_\textrm{int}) = \int_0^{T_\textrm{int}}  \big[W_\textrm{I}(t) I_{|k\rangle}(t) + W_\textrm{Q}(t)Q_{|k\rangle}(t)\big]dt,
\end{equation}
where we integrate the signal for a time $T_\textrm{int}$ long enough that all of the energy in the qubit readout pulse has decayed.  The weights $W_\textrm{I}(t)$ and $W_\textrm{Q}(t)$ are chosen to optimise the SNR of the qubit readout \cite{gambetta2007protocols}:
\begin{equation}
    W_\textrm{I}(t) = \langle I_{|e\rangle}(t) \rangle -\langle I_{|g\rangle}(t) \rangle
\end{equation}
\begin{equation}
    W_\textrm{Q}(t) = \langle Q_{|e\rangle}(t) \rangle -\langle Q_{|g\rangle}(t) \rangle.
\end{equation}
The variance of this integrated voltage is then given by
\begin{equation}
    \begin{split}
    \langle \Delta V_{|k\rangle}^2 (T_\textrm{int}) \rangle = \int_0^{T_\textrm{int}} \int_0^{T_\textrm{int}} \big[ W_\textrm{I}(t)W_\textrm{I}(t')\langle \hat{\zeta}_\textrm{I}(t) \hat{\zeta}_\textrm{I}(t') \rangle\big.\\ \big.+\ W_\textrm{Q}(t)W_\textrm{Q}(t')\langle \hat{\zeta}_\textrm{Q}(t) \hat{\zeta}_\textrm{Q}(t') \rangle \big]   dt\,dt'
    \end{split}
\end{equation}
Thus, we can define the total two-quadrature noise $N_\textrm{det}$ at the input of the ideal optical heterodyne detector (where any inefficiency in the real optical detector has been included in $\eta_\textrm{opt}$) in units of photons/s/Hz as
\begin{equation}
N_\textrm{det} = 1+N_\textrm{t},
\end{equation}
where $N_\textrm{t}$ is the noise added by the transducer,
\begin{equation}
\begin{split}
    N_\textrm{t} = \frac{\Gamma_\textrm{T}S_\textrm{t}(0)}{4G(T_\textrm{int})}\int_0^{T_\textrm{int}} \int_0^{T_\textrm{int}} \big[W_\textrm{I}(t)W_\textrm{I}(t')\big.\\+\ \big.W_\textrm{Q}(t)W_\textrm{Q}(t')\big]\,e^{-\Gamma_\textrm{T}|t-t'|/2}\,dt\,dt'+S_\textrm{b},
\end{split}
\end{equation}
and $G(T_\textrm{int}) = \frac{1}{2}\int_0^{T_\textrm{int}} \left[W_\textrm{I}(t)^2 + W_\textrm{Q}(t)^2\right]\,dt$.  

\section{State-space model}\label{sub:theory}
We can use the Heisenberg-Langevin equations obtained from Eq.~(3) in the Methods to model the response of the transducer to a pulse emitted by the circuit QED system.  Using the state-space model formalism the equations of motion can be written as
\begin{align}
    \mathbf{\dot{x}}(t) &= A\mathbf{x}(t) + B\mathbf{x}_\textrm{in}(t) \\
    \mathbf{x}_\textrm{out}(t) &= C\mathbf{x}(t) + D\mathbf{x}_\textrm{in}(t),
\end{align}
where $\mathbf{x} = (\hat{X}_1, \hat{Y}_1, \hat{Z}_1, \hat{X}_2, \hat{Y}_2, \hat{Z}_2)^T$ is a vector of quadrature amplitudes, with $\hat{X}_l$, $\hat{Y}_l$ and  $\hat{Z}_l$ ($l={1,2}$) corresponding to the dimensionless optical, microwave and mechanical mode quadratures respectively.  The optical quadratures can be defined in terms of creation/annihilation operators as $\hat{X}_1 = \frac{1}{2}(\hat{a}^\dagger + \hat{a})$ and $\hat{X}_2 = \frac{i}{2}(\hat{a}^\dagger-\hat{a})$, with the microwave and mechanical quadratures defined analogously.  The input and output fields may also be written in vector form:
\begin{widetext}
\begin{gather}
\mathbf{x}_{\textrm{in}} = ( \hat{X}_\textrm{1,in}, \hat{X}_\textrm{1,in,int},\hat{Y}_\textrm{1,in}, \hat{Y}_\textrm{1,in,int}, \hat{Z}_\textrm{1,in,int}, \hat{X}_\textrm{2,in}, \hat{X}_\textrm{2,in,int},\hat{Y}_\textrm{2,in}, \hat{Y}_\textrm{2,in,int}, \hat{Z}_\textrm{2,in,int})^T \\
\mathbf{x_\textrm{out}} = ( \hat{X}_\textrm{1,out},\hat{Y}_\textrm{1,out}, \hat{X}_\textrm{2,out},\hat{Y}_\textrm{2,out})^T,
\end{gather}
\end{widetext}
The transducer can be viewed as a phase-preserving amplifier with near unity gain \cite{caves1982quantum}, so the equations of motion are independent of the phases of the optical and microwave pumps.  Thus under an appropriate choice of these arbitrary phases, the elements of the matrices A, B, C, D in the state-space model can all be made real.  We can transform to a rotating frame that removes the free evolution of the quadratures and describe the state-space model with the following matrices:
\begin{widetext}
\begin{equation}
A_\textrm{RWA} =
    \begin{pmatrix}
      -\frac{\kappa_\textrm{o}}{2} & 0 &0 & 0 & 0 & -g_\textrm{o}\bar{a}\\
      0 & -\frac{\kappa_\textrm{e}}{2} & 0 & 0 & 0 & -g_\textrm{e}\bar{b} \\
      0 & 0 & -\frac{\gamma_\textrm{m}}{2} & -g_\textrm{o}\bar{a} &
      -g_\textrm{e}\bar{b} & 0 \\
      0 & 0 & g_\textrm{o}\bar{a} & -\frac{\kappa_\textrm{o}}{2} & 0 & 0 \\
      0 & 0 & g_\textrm{e}\bar{b} & 0 & -\frac{\kappa_\textrm{e}}{2} & 0 \\ 
      g_\textrm{o}\bar{a} & g_\textrm{e}\bar{b} & 0 & 0 & 0 & -\frac{\gamma_\textrm{m}}{2} \\  
    \end{pmatrix},
\end{equation}

\begin{equation}
A_\textrm{counter}=
    \begin{pmatrix}
      0 & 0 & -g_\textrm{o}\bar{a}\sin(2\omega_\textrm{m}t) & 0 & 0 & g_\textrm{o}\bar{a}\cos(2\omega_\textrm{m}t)\\
      0 & 0 & -g_\textrm{e}\bar{b}\sin(2\omega_\textrm{m}t) & 0 & 0 & g_\textrm{e}\bar{b}\cos(2\omega_\textrm{m}t) \\
      -g_\textrm{o}\bar{a}\sin(2\omega_\textrm{m}t) & -g_\textrm{e}\bar{b}\sin(2\omega_\textrm{m}t) & 0 & g_\textrm{o}\bar{a}\cos(2\omega_\textrm{m}t) &
      g_\textrm{e}\bar{b}\cos(2\omega_\textrm{m}t) & 0 \\
      0 & 0 & g_\textrm{o}\bar{a}\cos(2\omega_\textrm{m}t) &0 & 0 & g_\textrm{o}\bar{a}\sin(2\omega_\textrm{m}t) \\
      0 & 0 & g_\textrm{e}\bar{b}\cos(2\omega_\textrm{m}t) &0 & 0 & g_\textrm{e}\bar{b}\sin(2\omega_\textrm{m}t) \\
      g_\textrm{o}\bar{a}\cos(2\omega_\textrm{m}t) & g_\textrm{e}\bar{b}\cos(2\omega_\textrm{m}t) & 0 & g_\textrm{o}\bar{a}\sin(2\omega_\textrm{m}t) & g_\textrm{e}\bar{b}\sin(2\omega_\textrm{m}t) & 0 \\  
    \end{pmatrix},
\end{equation}
\begin{equation}
A = A_\textrm{counter} + A_\textrm{RWA},  
\end{equation}
\begin{equation}
    B =    
    \begin{pmatrix}
      M & 0\\
      0 & M
    \end{pmatrix}
\end{equation}
\begin{equation}
    M = 
    \begin{pmatrix}
      \sqrt{\kappa_\textrm{o,ext}}& \sqrt{\kappa_\textrm{o,int}} & 0&0&0 \\
      0&0&\sqrt{\kappa_\textrm{e,ext}}&\sqrt{\kappa_\textrm{e,int}}&0 \\
       0 & 0 & 0 & 0 & \sqrt{\gamma_\textrm{m}}
    \end{pmatrix},
\end{equation}
\begin{equation}
    C = 
    \begin{pmatrix}
      \sqrt{\kappa_\textrm{o,ext}} &0&0&0&0&0 \\
      0&\sqrt{\kappa_\textrm{e,ext}}&0&0&0&0 \\
      0&0&0&\sqrt{\kappa_\textrm{o,ext}}&0&0 \\
      0&0&0&0&\sqrt{\kappa_\textrm{e,ext}}&0
    \end{pmatrix},
\end{equation}
\setcounter{MaxMatrixCols}{20}
\begin{equation}
    D = 
    \begin{pmatrix}
      -1 &0 &0 &0 &0 &0 &0 &0 &0 &0 \\
      0 &0 &-1 &0 &0 &0 &0 &0 &0 &0\\
      0 &0 &0 &0 &0 &-1 &0 &0 &0 &0\\
      0 &0 &0 &0 &0 &0 &0 &-1 &0 &0
    \end{pmatrix}.
\end{equation}
\end{widetext}
%TC:ignore
%\bibliographyMethods{methodsReferences}


%\bibliography{references}
\FloatBarrier
\setcounter{figure}{0}
\renewcommand{\figurename}{\textbf{Supplementary Fig.}}

%\begin{figure*}[t]
%    \includegraphics{detailedMeasurementChainV3.eps}
%    \caption{\textbf{Experimental Schematic}  (a) Microwave layout demonstrating the exact configuration of the qubit readout/control pulses and the pumps for the electro-optic transducer.  (b) Legend for various different microwave and optical components.  (c) Cryogenic portion of the experiment.  (d) Demodulation and detection scheme.  The two digitisers allow for simultaneous measurement of the microwave and optical signals emitted from the transducer.  (e) Simplified schematic of optical beam layout and balanced heterodyne detector.}
%    \label{fig:detailedDiagram}
%\end{figure*}
%TC:endignore